\documentclass[fleqn,usenatbib]{mnras}

\usepackage{newtxtext,newtxmath}

\usepackage[T1]{fontenc}
\usepackage{ae,aecompl}

\usepackage{graphicx}	
\usepackage{amsmath}	
\usepackage{amssymb}	

\title[Cosmic rays in starburst nuclei]{Cosmic ray transport and radiative processes in nuclei of starburst galaxies}

\author[E. Peretti et al.]{Enrico Peretti$^{1,2}$ \thanks{E-mail: enrico.peretti@gssi.it},
Pasquale Blasi$^{1,2}$ \thanks{E-mail: pasquale.blasi@gssi.it},
Felix Aharonian$^{1,3,4}$ \thanks{E-mail: felix.aharonian@mpi-hd.mpg.de} and \newauthor Giovanni Morlino$^{1,5}$ \thanks{E-mail: giovanni.morlino@gssi.it}
\\
$^{1}$Gran Sasso Science Institute, Viale F. Crispi 7, L'Aquila, Italy \\
$^{2}$INFN/Laboratori Nazionali del Gran Sasso, via G. Acitelli 22, Assergi (AQ), Italy \\
$^{3}$Max-Planck-Institute f{\"u}r Kernphysik, Saupfercheckweg 1, D-69117 Heidelberg, Germany \\
$^{4}$Dublin Institute for Advanced Studies, 31 Fitzwilliam Place, Dublin 2, Ireland \\
$^{5}$INAF/Osservatorio Astrofisico di Arcetri, L.go E. Fermi 5, Firenze, Italy \\
}

\date{Accepted XXX. Received YYY; in original form ZZZ}

\pubyear{2018}

\begin{document}
\label{firstpage}
\pagerange{\pageref{firstpage}--\pageref{lastpage}}
\maketitle

\begin{abstract}
The high rate of star formation and supernova explosions of starburst galaxies make them interesting sources of high energy radiation. Depending upon the level of turbulence present in their interstellar medium, the bulk of cosmic rays produced inside starburst galaxies may lose most of their energy before escaping, thereby making these sources behave as calorimeters, at least up to some maximum energy. Contrary to previous studies, here we investigate in detail the conditions under which cosmic ray confinement may be effective for electrons and nuclei and we study the implications of cosmic ray confinement in terms of multifrequency emission from starburst nuclei and production of high energy neutrinos. The general predictions are then specialized to three cases of active starbursts, namely M82, NGC253 and Arp220. Both primary and secondary electrons, as well as electron-positron pairs produced by gamma ray absorption inside starburst galaxies are taken into account. Electrons and positrons produced as secondary products of hadronic interactions are found to be responsible for most of the emission of leptonic origin. In particular, synchrotron emission of very high energy secondary electrons produces an extended emission of hard X-rays that represent a very interesting signature of hadronic process in starburst galaxies, potentially accessible to current and future observations in the X-ray band. A careful understanding of both the production and absorption of gamma rays in starburst galaxies is instrumental to the assessment of the role of these astrophysical sources as sources of high energy astrophysical neutrinos. 
\end{abstract}




\section{Introduction}

Starburst galaxies (SBGs) are unique sources showing a very intense star formation activity, at a level that can be as high as $\dot{M} \sim 10 \div 100$ $M_{\odot} yr^{-1}$, as discussed by \citet{2004ApJ...606..271G}. Their star forming regions, called starburst nuclei (SBNi), typically extend on few hundred parsec and are often observed in the cores of SGBs. The rapid star forming activity, which reflects in an enhanced far infrared (FIR) luminosity \cite[]{2003A&A...401..519M}, leads to a correspondingly higher supernova rate, $\mathcal{R}_{SN} \sim 0.1 \div 1 \; yr^{-1}$, thereby suggesting that SBNi may be efficient sites of cosmic ray (CR) production.

The density of interstellar medium (ISM) in SBNi is estimated to be of the order of $n_{ISM} \sim 10^2$ $cm^{-3}$, with a mass in the form of molecular clouds $M_{mol} \sim 10^8 M_{\odot}$. The mass in the form of  ionized gas is typically a few percent of that of the neutral gas \citep[a detailed discussion for the case of M82 was presented by][]{2001ApJ...552..544F}. The FIR radiation can easily reach an energy density of $U_{RAD} \sim 10^3$ $ eV/cm^3$ while the strength of the inferred magnetic field is of order $B \sim 10^2 \div 10^3$ $\mu G$ \citep[e.g.][]{2006ApJ...645..186T}. Moreover, the high supernova rate, together with a possible coexisting AGN activity, are expected to highly perturb the global SBN environment. Strong winds are in fact observed in many starbursts at every wavelength with estimated velocities of several hundred kilometers per seconds as reported for the case of M82 by \citet{2009ApJ...697.2030S}, \citet{1538-4357-642-2-L127} and \citet{1991ApJ...369..320S}.

Winds and turbulence play a fundamental role in CR transport in SBNi. The former lead to advection of CRs, a phenomenon that typically acts in the same way for CRs of any energy. The latter is responsible for CR diffusion through resonant scattering off perturbations in the magnetic field. The combination of wind advection, diffusion and energy losses shapes the transport of CRs in SBNi and determines whether or not the bulk of CRs is confined inside the nucleus, namely if particles lose most of their energy before escaping the nucleus (through either advection or diffusion). The phenomenon of CR confinement is crucial to understand the production of non-thermal radiation and neutrinos in SBGs. At energies where losses act faster than escape, the production of secondary electrons and positrons is prominent and in fact secondary electrons can be shown to be dominant upon primary electrons, for typical values of environmental parameters. In turn this implies that  secondary electrons shape the multifrequency emission of SBNi through their synchrotron (SYN) and inverse Compton (IC) emission, a situation quite unlike the one of our Milky Way. Here we study in detail under which conditions SBNi behave as calorimeters: we find that for the conditions expected in SBNi, transport is dominated by advection with the wind up to very high energies. At sufficiently high energies (depending upon the level of turbulence), diffusion starts being dominant and leads to a transition to a regime where CR protons can leave the SBN before appreciable losses occur. In passing, we notice that the wind itself has been proposed as possible site where particle acceleration to extremely high energies might take place \cite[]{Anchordoqui_1999_Wind_1,Romero_wind,2018PhRvD..97f3010A}.

Several models have been previously developed to describe the behaviour of CRs in starburst environments and infer their high energy emission \cite[]{1996ApJ...460..295P,2004ApJ...617..966T,2008A&A...486..143P,2010MNRAS.401..473R,0004-637X-762-1-29}. In all these works diffusion effects were typically accounted for by assuming a diffusive escape time defined by a power law energy dependence with slope $\delta = 0.5$ and a normalization of few millions of years at GeV energies. On the other hand, \cite{Yoast-Hull:2013wwa} assumed that CR transport \textbf{is} dominated solely by wind advection and energy losses, while diffusion would be negligible.  \cite{2018MNRAS.474.4073W} focused on hadronic gamma-ray emission in the framework in which SBNi are treated as calorimeters, whereas \citet{Sudoh:2018ana} modeled the proton transport accounting for wind advection and Kolmogorov-like diffusion. SBGs have been also discussed as possible neutrino factories, both as isolated sources \cite[]{2003ApJ...586L..33R,2009ApJ...698.1054D,2015PhDT........94T} and as possible relevant contributors to the global diffuse flux \cite[]{Loeb:2006tw,2011ApJ...734..107L,Tamborra:2014xia,Bechtol:2015uqb}. 

In this article we improve with respect to previous studies in several respects: 1) the issue of calorimetric behaviour of SBNi is addressed in a quantitative way, by discussing how different assumptions about the turbulence in the ISM of SBNi changes the the escape of CRs from the confinement volume as compared with the role of an advecting wind. This means that we can now also describe the transition from calorimetric behaviour to diffusion dominated regime. This transition reflects into features in the spectrum of high energy gamma rays from the decay of neutral pions. 2) The spectrum of secondary electrons is self-consistently calculated taking into account advection, diffusion and energy losses, so as to have at our disposal a self-consistent calculation of the multifrequency spectrum of radiation produced by electrons (primary and secondary) through SYN and ICS. 3) The absorption of gamma rays as due to electron-positron pair production inside the starburst region is taken into account. This allows us to determine the spectrum of gamma rays reaching us from an individual SBG and the contribution to the diffuse gamma ray background. 4) The secondary electrons resulting from the decay of charged pions and from absorption of gamma rays on the photon background inside a SBN both contribute to the production of a diffuse X-ray radiation as due to SYN emission. The detection of such emission would represent an unambiguous signature of the calorimetric behaviour of SBNi. 


The paper is organized as follows: in \S \ref{Sezione_2} we describe the theoretical approach for the calculation of the CR distribution function inside a generic SBN and the associated photon and neutrino spectra. In \S \ref{Sezione_3} we discuss how different assumptions on the diffusion coefficient affects the confinement of cosmic rays inside SBNi and in \S \ref{Sezione_4} we apply our model to three SBGs, namely NGC253, M82 and Arp220 so as to have a calibration of our calculation to their observed multifrequency spectra. This allows us to have a physical understanding of CR transport in a SBG that can be applied to the determination of SBGs to the diffuse gamma and neutrino emission that will be discussed in detail in a forthcoming paper. We draw our conclusions in \S \ref{Conclusioni}.

\section{Cosmic ray transport in a SBN}
\label{Sezione_2}

Since the starburst nucleus of a SBG is rather compact and populated by both gas and sources, the simplest approach to CR transport in such a region is represented by a leaky-box-like model in which the injection of CR protons and electrons is balanced by energy losses, advection with a wind and diffusion:
\begin{equation} 
\phantom{xxxxxxxxx} \frac{f(p)}{\tau_{\rm loss}(p)} + \frac{f(p)}{\tau_{\rm adv}(p)} + \frac{f(p)}{\tau_{\rm diff}(p)} = Q(p) ,
 \label{eq:CR_Equation} 
\end{equation}
where $f$ is the CR distribution function, $Q$ is the injection term due to supernovae explosions, while $\tau_{\rm loss}$, $\tau_{\rm adv}$ and $\tau_{\rm diff}$ are the timescales of energy losses, wind advection and diffusion, respectively. The characteristic time for energy losses is derived combining effects due to radiative emission and collisions, namely
\begin{equation}
\phantom{xxxxxxxxxxxxxxx} \frac{1}{\tau_{\rm loss}}=  \sum_i \left( - \frac{1}{E} \frac{dE}{dt} \right)_i,
\label{CR_losses_timescale}
\end{equation}
where $i$ sums over ionization, proton-proton collisions and Coulomb interactions in the case of protons, whereas in the case of electrons it represents losses due to ionization, synchrotron, inverse Compton and bremsstrahlung. The detailed expressions adopted for each channel are reported for completeness in Appendix \ref{app:timescales}. The advection timescale $\tau_{\rm adv}$ is the ratio between the SBN size and the wind speed, i.e. $\tau_{\rm adv}=R/v_{\rm wind}$, and provides an estimate of the typical time in which particles are advected away from the SBN. Similarly, the diffusion timescale is taken as $\tau_{\rm diff}(p)=R^2/D(p)$, where $D(p)$ is the diffusion coefficient as a function of particle momentum.  Here we adopt an expression for $D(p)$ that is inspired by the quasi-linear formalism
\begin{equation}
\phantom{xxxxxxxxxxxxxxxx} D(p)=\frac{r_L(p)v(p)}{3\mathcal{F}(k)},
\label{Kolmogorov}
\end{equation} 
where $\mathcal{F}(k)=kW(k)$ is the normalized energy density per unit logarithmic wavenumber $k$, and $W(k)=W_0(k/k_0)^{-d}$, with $k_0^{-1}=L_0$ characteristic length scale at which the turbulence is injected. We calculate $\mathcal{F}(k)$ by requiring the following normalization condition
\begin{equation}
\phantom{xxxxxxxxxxxxxxx}  \int^{\infty}_{k_0}W(k)dk = \left( \frac{\delta B}{B} \right)^2 = \eta_{B}.
\end{equation}
In order to bracket plausible models of CR diffusion in SBNi we adopt three models of diffusion: 1) a benchmark model in which $d=5/3$, $L_{0}=1$ pc and $\eta_B=1$ (model A), which leads to a Kolmogorov-like diffusion coefficient with asymptotic energy dependence $\sim E^{1/3}$ ; 2) a case in which $d=0$ and $\eta_B=1$ which leads to a Bohm diffusion coefficient (Model B); 3) a case in which $d=5/3$ and $\eta_B$ is normalized in such a way that the diffusion coefficient at 10 GeV is $\sim 3\times 10^{28}~\rm cm^{2}/s$, which is supposed to mimic the diffusion coefficient inferred for our Galaxy (Model C). The latter case is expected to lead to faster diffusion and lesser confinement of CR protons in the SBN. In Model A, the choice of $L_{0}\ll R$ was made to mimic an ISM with strong turbulence on pc scales. For the cases above we choose a magnetic field $B=200\mu G$ and a size of the SBN $R=200$ pc.

Given the starburst nature of the sources, it is expected that the main injection of CRs in SBNi occurs through supernova explosions. The injection term $Q$ in Eq.\ref{eq:CR_Equation} is assumed to be constant in the entire spherical volume and is computed as
\begin{equation}
 \phantom{xxxxxxxxxxxxxxxx} Q(p)= \frac{\mathcal{R}_{\rm SN}\mathcal{N}_{p}(p)}{V},
\label{CR_Injection}
\end{equation}
where $\mathcal{N}_{p}(p)$ is the injection spectrum of protons from an individual SNR, and $\mathcal{R}_{\rm SN}$ is the rate of SN explosions in the SBN volume $V$. Assuming that the spectrum of accelerated CR protons has the shape of a power law in momentum with index $\alpha$ up to a maximal value $p_{max,p}$, we can write
\begin{equation}
\phantom{xxxxxxxxxxx}    \mathcal{N}_{p}(p) \propto  \left( \frac{p}{m_p c} \right)^{- \alpha}e^{-p/p_{p,\max}},
\end{equation}
where the normalization constant is calculated by requiring that 
\begin{equation}
 \phantom{xxxxxxxxxxx} \int^{\infty}_0 4 \pi p^2 \mathcal{N}_{p}(p)T(p)  dp = \xi_{\rm CR} E_{\rm SN},
\end{equation}
with $T(p)$ the kinetic energy of particles, $\xi_{\rm CR}$ the acceleration efficiency (of order $10\%$), and $E_{\rm SN}$ the explosion energy for which we adopt the typical value of $10^{51}$ erg.

For electrons, the slope of the injection spectrum is assumed to be the same as for protons, but the cutoff is assumed to be as found in calculations of diffusive shock acceleration in the presence of energy losses and Bohm diffusion \cite[]{2007A&A...465..695Z,Blasi:2009ix}:
\begin{equation}
\phantom{xxxxxxxxxxx}    \mathcal{N}_{e}(p) \propto  p^{- \alpha}e^{-(p/p_{e,max})^{2}}.
\end{equation}
Throughout the paper we assume that $p_{p,\max}=10^{5}$ TeV/c and $p_{e,\max}=10$ TeV/c. We also assume that the spectrum of injected electrons has a lower normalization than protons by a factor $\sim 50$, as also assumed by \cite{2004ApJ...617..966T} and \citet{Yoast-Hull:2013wwa} and close to what is inferred for our Galaxy.


In order to quantify the confinement properties of SBNi, namely the situations in which CR protons and electrons lose energy before escaping the SBN, we adopt some reference values for the parameters,  summarized in Table~\ref{tab:parameters_table} and adopted in the estimates of time scales for the different processes. We refer to this set of parameters as our ``\textit{reference case}''.
\begin{table}
\centering
\begin{tabular}{|l|r|}
\hline
 List of parameters  &  Value  \\ \hline \hline
 $D_L$ $({\rm Mpc})$ $[\rm redshift]$ & $3.8 \; [8.8 \times 10^{-4}]$   \\ \hline
 $\mathcal{R}_{\rm SN}$ $({\rm yr^{-1}})$ & $0.05$  \\ \hline
 $R$ (pc) & $200$    \\ \hline
 $\alpha$  & $4.25$   \\ \hline
 $B$ $(\mu$G) & $200$   \\ \hline
 $v_{\rm wind}$ (km/s) & $500$   \\ \hline
 $M_{\rm mol}$ $(10^8 M_{\odot})$ & $1.0$   \\ \hline 
 $n_{\rm ISM}$ $\rm (cm^{-3})$ & $125$   \\ \hline
 $n_{\rm ion}$ $\rm (cm^{-3})$ & $18.75$   \\ \hline
 $T_{\rm plasma} \rm (K)$  & $6000$   \\ \hline
 $U^{\rm FIR}_{\rm Rad}$ ($\rm eV \,cm^{-3}$) [kT (meV)] & $1101$ $[3.5]$  \\ \hline
 $U^{\rm MIR}_{\rm Rad}$ ($\rm eV \,cm^{-3}$) [kT (meV)] & $330$ $[8.75]$  \\ \hline
 $U^{\rm NIR}_{\rm Rad}$ ($\rm eV \,cm^{-3}$) [kT (meV)] & $330$ $[29.75]$ \\ \hline
 $U^{\rm OPT}_{\rm Rad}$ ($\rm eV \,cm^{-3}$) [kT (meV)] & $1652$ $[332.5]$  \\ \hline
\end{tabular}
\caption{\label{tab:parameters_table} Table of parameters for the adopted reference case. $D_L$ is the luminosity distance of the source, $\mathcal{R}_{\rm SN}$ is the supernova rate, $R$ is the radius of the SBN, $\alpha$ is the injection index in momentum, $B$ is the mean magnetic field and $v_{\rm wind}$ is the outgoing wind velocity. The molecular cloud mass in the SBN is represented by $M_{\rm mol}$ and it coincides with an overall particle density given by $n_{\rm ISM}$. The ionized gas density is expressed by $n_{\rm ion}$ which has a temperature $T_{\rm plasma}$. The last four lines show the energy density $U$ and the temperature $kT$ of the three IR components due to dust and the optical one due to stars.}
\end{table}

\begin{figure}
\centering
\includegraphics[width=0.45\textwidth]{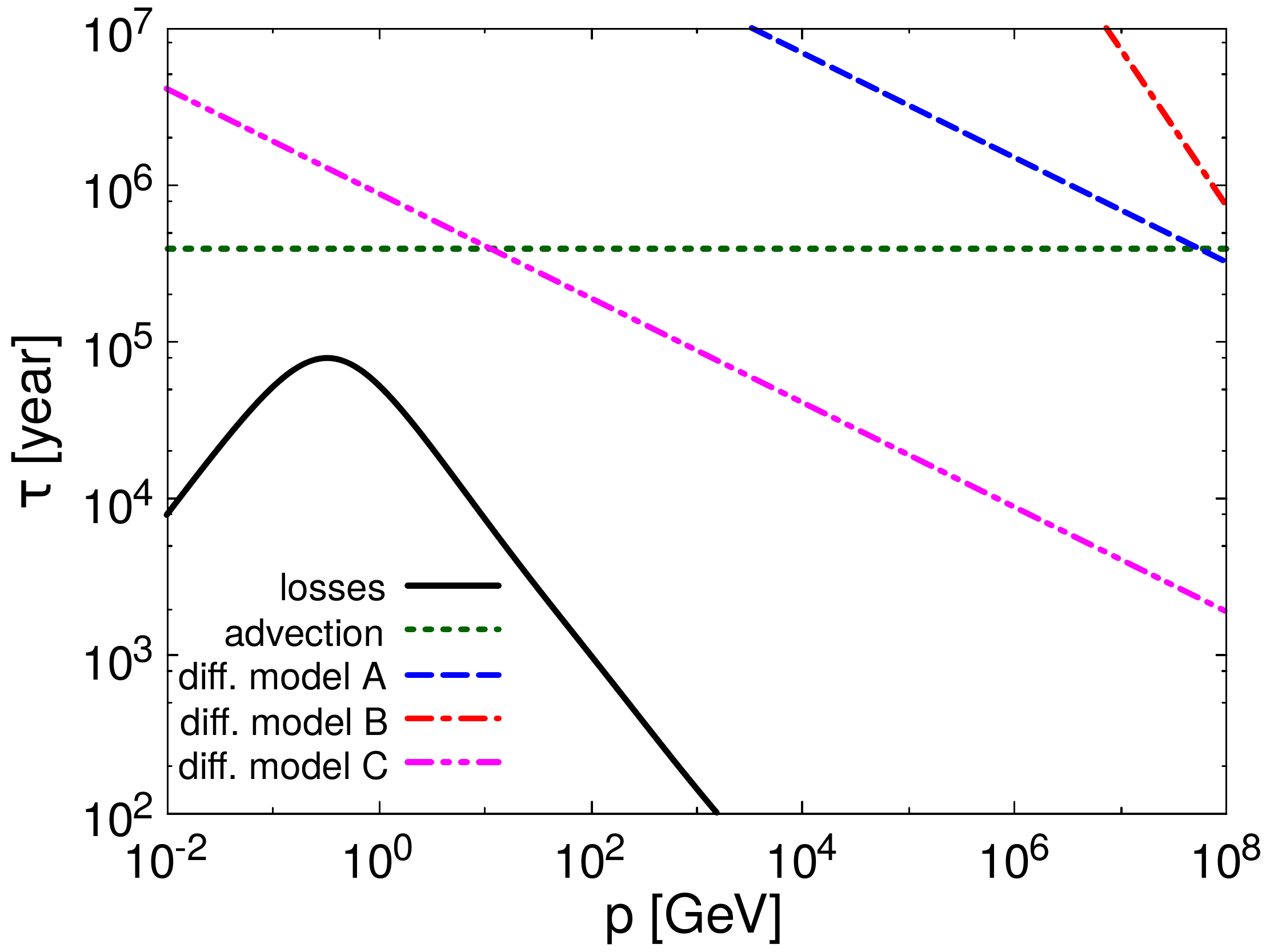}\quad\includegraphics[width=0.45\textwidth]{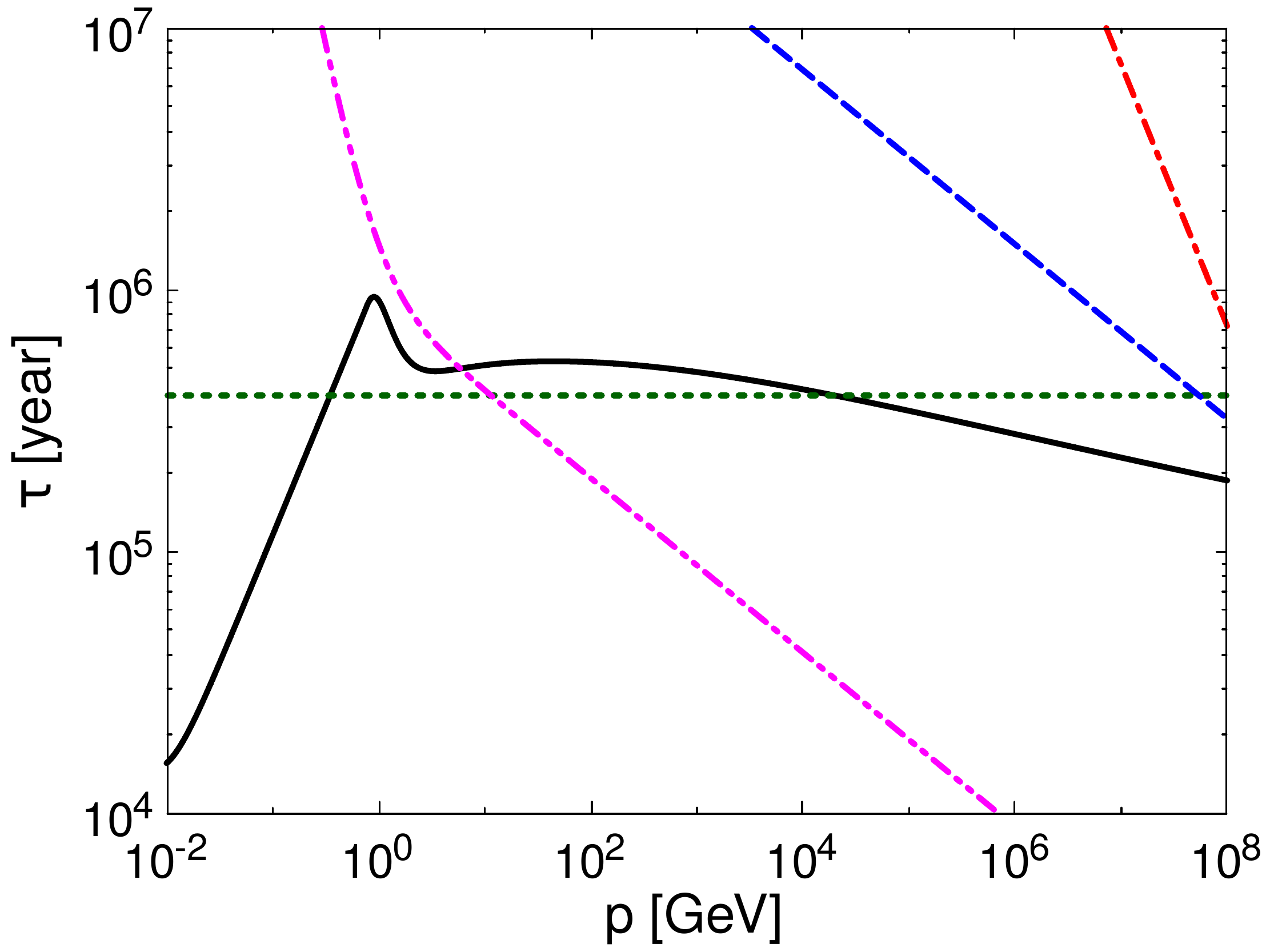}
\caption{\label{fig:TimescalesPrimo} Energy dependence of the characteristic timescales (expressed in years) of cosmic ray electrons (upper panel) and protons (lower panel) for the parameters of our reference case. Black thick lines represent energy losses, green dotted lines show the advection timescales. The timescales of diffusion are represented by blue dashed lines in the case of Kolmogorov, red dot-dashed in the case of Bohm and magenta dot-dot-dashed in the MW-like case.}
\end{figure} 

The time scales for diffusion, advection and energy losses for CR electrons and protons are shown in the top and bottom panels of Figure \ref{fig:TimescalesPrimo}, respectively. The horizontal (dotted green) line refers to the advection time scales, which is clearly independent of energy and is the same for electrons and protons. For typical values of the radius $R \sim 10^2$ $pc$ \citep[see i.e.][]{1538-4357-576-1-L19} and wind velocity $v_{\rm wind} \sim 10^2 \div 10^3$ km s$^{-1}$ \citep[see e.g.][]{1998ApJ...493..129S}, the advection timescale is of the order of a few hundred thousand years. 

The time scale for losses of electrons (solid black line) shows an increasing trend for low momenta, reflecting the dominant ionization and bremsstrahlung channels. At high energy synchrotron and inverse Compton scattering start being important and the loss time drops with energy approximately as $E^{-1}$. The time scale for diffusive escape from the SBN for Model A (dashed blue line), Model B (dash-dotted red line) and Model C (dash-dot-dotted magenta line) are also shown. For all these models it is clear that energy losses dominate the transport of electrons at all energies. For Models A and B, the escape of electrons occurs due to wind advection, while for Model C there is a transition from advection to diffusion at energies $\sim$GeV. In any case, SNBi behave as electron calorimeters.

For CR protons, energy losses are dominated by ionization at low energies and by inelastic pp collisions at high energy. For the Models A and B of diffusive transport, the loss time scales is always shorter than the time for diffusive escape. However, transport is dominated by wind advection at all energies of interest. The time scale for advection and pp scattering remain comparable over many orders of magnitude in energy, due to the fact that both are roughly energy independent. In other words, SBNi behave as approximate, though not perfect, calorimeters. In Model C, CR transport is dominated by diffusion for energies above $\sim$ GeV, and only a small fraction of the energy is lost during propagation. This latter case does not appear to be well motivated and is shown here only as a rather extreme scenario. Moreover, as we discuss below, the multifrequency spectra of individual SBGs are not easy to explain in the context of Model C.

\subsection{Secondary and tertiary electrons and neutrinos}

Electron-positron pairs are copiously produced in SBNi because of the severe rate of energy losses of CR protons. Following the approach put forward by \citet{2006PhRvD..74c4018K}, we compute the pion injection rate as
\begin{equation}
 \phantom{xxxx} q_{\pi}(E_{\pi}) = \frac{c n_{\rm ISM}}{K_{\pi}} \sigma_{pp} 
 		\left(m_p c^2+\frac{E_{\pi}}{K_{\pi}} \right) n_p \left(m_p c^2+\frac{E_{\pi}}{K_{\pi}}\right),
\label{pion_injection}
\end{equation}
where $K_{\pi}\sim 0.17$ is the fraction of kinetic energy transferred from the parent proton to the single pion. $n_p(E)$ is the proton distribution function in energy, which is linked to the distribution in momentum by $n_p(E)dE= 4 \pi p^2 f_p(p) dp$. The secondary electron injection (here we refer to electrons as the sum of secondary electrons and positrons) is then computed as follows:
\begin{equation}
 \phantom{xxxxxxxxxx}  q_e(E_e)= 2 \int_{E_e}^{\infty} q_{\pi}(E_{\pi}) \tilde{f}_e \left(\frac{E_{e}}{E_{\pi}} \right) \frac{dE_{\pi}}{E_{\pi}},
\end{equation}
where $\tilde{f}_e$, defined in equations (36-39) of \citet{2006PhRvD..74c4018K}, is reported in Appendix \ref{app:secondaries}. As we discuss below, gamma rays are also produced as a result of the production and decay of neutral pions. 

As illustrated in Table \ref{tab:parameters_table}, the density of FIR photons is large enough that the opacity for photons above threshold for pair production is $\tau_{\gamma\gamma}\gg 1$ (see discussion in Appendix \ref{Appendice_Stime_Analitiche}), so that photons with $E_{\gamma} \gtrsim 10$ TeV are absorbed inside the SBN, and give rise to $e^{\pm}$ pairs that we refer to as {\it tertiary electrons}.


The rate of injection of tertiary electrons is calculated using the leading particle approximation suggested by \citet{2013SAAS...40.....A}. The corresponding spectrum of injected pairs is
\begin{align}
\begin{split}
 \phantom{xxxxxxx} q_{e}(E,r) & =  \int d\epsilon \; n_{\rm bkg}(\epsilon) n_{\gamma}(E,r) \sigma_{\gamma \gamma}(E,\epsilon) c \\
& = n_{\gamma}(E,r)c \tau_{\gamma \gamma}(E)/R,
\end{split}
\label{pair_production_inj}
\end{align}
where $n_{\rm bkg}(\epsilon)$ is the target background photon density and $n_{\gamma}(E)$ is the gamma-ray photon density, related to the photon emissivity through the expression $\epsilon_{\gamma}(E,r) \approx n_{\gamma} (E,r) c/(4 \pi R) $, which accounts for $\pi_0$ decay, synchrotron, inverse Compton and bremsstrahlung emission of electrons. All these radiation mechanisms are discussed in the following subsection.

The $p \gamma$ interaction could also provide a contribution to secondary electrons, provided the maximum energy of CR protons is higher than $\sim 1.5 \times 10^{8}$ GeV, a case that we do not consider here, but could retain some interest in other contexts. 

The equilibrium spectrum of secondary (and tertiary) electrons is calculated by solving Eq. \ref{eq:CR_Equation}. However , since for electrons energy losses are always dominant, the equilibrium spectrum is well approximated by $f_{{\rm sec},e}(p)= q_e(p) \tau_{\rm loss}(p)$. Such approximation is also valid for tertiary electrons above the production threshold. Nevertheless, below such threshold the spectrum is not vanishing but is populated by electrons that lose energy during the propagation. To account also for this component we calculate the spectrum of tertiary electrons as
\begin{equation}
  f_{{\rm ter},e}(E)= \frac{\tau_{\rm loss}(E)}{E} \int_{E}^{\infty} E' q_e(E') dE'
\end{equation}
where $q_e$ is taken from Eq.~(\ref{pair_production_inj}).

We also computed the production rate of neutrinos from $pp$ interactions, following the approach proposed by \cite{2006PhRvD..74c4018K}, where the muon neutrino injection was written as
\begin{equation}
 \phantom{xxxxxxx}  q_{\nu_{\mu}}(E)= 2 \int_{0}^{1} \left[ f_{\nu_{\mu}^{(1)}}(x)+f_{\nu_{\mu}^{(2)}}(x) \right] 
 				q_{\pi} \left( \frac{E}{x} \right) \frac{dx}{x},
\label{neutrino_formula}
\end{equation}
with $x= E/E_{\pi}$ and the functions $f_{\nu_{\mu}^{(1)}}$ and $f_{\nu_{\mu}^{(2)}}$, as reported in Appendix \ref{app:secondaries}, describe muon neutrinos produced by the direct decay $\pi \longrightarrow \mu \nu_{\mu}$ and by the muon decay $\mu \longrightarrow \nu_{\mu} \nu_e e$, respectively. The latter process also produces electron neutrinos which are described by the same equation \ref{neutrino_formula} where the square bracket is replaced with the function $f_{\nu_{e}}$ (see Appendix \ref{app:secondaries}). During propagation over cosmological distances, neutrino oscillations lead to equal distribution of the flux among the three flavors.  


\subsection{Non thermal radiation from SBNi}

Neutral pion decay is the leading process for the production of $\gamma$-rays in SBNi. Following the approach of \cite{2006PhRvD..74c4018K}, we calculate the photon emissivity in the following way
\begin{equation}
 \phantom{xxxxxxxxx} 4 \pi \; \epsilon_{\gamma}(E)=   2 \int_{E_{\min}}^{\infty} \frac{q_{\pi}(E_{\pi})}{\sqrt{E_{\pi}^2 - m_{\pi}^2c^4}} dE_{\pi} ,
\end{equation} 
where $E_{\min}=E+m_{\pi}^2c^4/(4E) $ and $q_{\pi}$ is defined in equation \ref{pion_injection}.

The emissivity due to bremsstrahlung is calculated here following \cite[]{1971NASSP.249.....S}:
\begin{equation}
 \phantom{xxxxxxx} 4 \pi \; \epsilon_{\rm brem}(E)=   \frac{n_{\rm ISM} \sigma_{\rm brem} c }{E} \int_{E}^{\infty} N_e(E_e,r) dE_e,
\end{equation}
where $\sigma_{\rm brem} \approx 3.4 \times 10^{-26} \rm cm^2$.

The synchrotron emissivity is calculated using the simplified approach proposed by \cite{2013LNP...873.....G}, namely assuming that all energy is radiated at the critical frequency, $\nu_{\rm syn}=\gamma^2 eB/2\pi m_e c$:
\begin{align}
\begin{split}
    \phantom{xxxxxxxxxxxx}  4 \pi \;  \epsilon_{\rm syn}(\nu) d \nu =  P_{\rm syn}(\gamma) N_e(\gamma) d \gamma   \\ \gamma= \sqrt{\frac{\nu}{\nu_{\rm syn}}} \; \; \frac{d \gamma}{d \nu}= \frac{\nu^{-1/2}}{2 \nu_{\rm syn}^{1/2}},
\end{split}
\label{SY_IC}
\end{align}
where $P_{\rm syn}$ is the total power emitted by a single electron \cite[]{1986rpa..book.....R, 2011hea..book.....L} (see Appendix \ref{app:timescales}). 

The low energy background thermal radiation plays a very important role both as a target for ICS and for $\gamma \gamma$ absorption and pair production. We model the dust thermal contribution in the FIR domain with a diluted blackbody (DBB) as proposed by \citet{0004-637X-568-1-88} and \citet{2008A&A...486..143P}, or possibly a combination of them in order to model different kinds of dust emitting at different temperatures. The single-temperature DBB has the following expression
\begin{equation}
\label{Diluted_BB}
\phantom{xxxxxxxxx} n_{\rm FIR}(E)= C_{\rm dil} \frac{8 \pi}{(hc)^3} \frac{E^2}{\exp^{E/kT}-1} \left( \frac{E}{E_0} \right)^{\sigma}. 
\end{equation}
This functional shape allows the dust spectrum to be a pure black body above the energy $E_0$, whereas at lower energies it reduces to a grey body spectrum $\propto E^{2+\sigma}$, where the dust spectral index $\sigma$ generally assuming values between $0$ and $2$ \citep[see][]{0004-637X-568-1-88}. The normalization $C_{\rm dil}$ is obtained from a fit to the IR spectrum of SBGs, while the stellar contribution, treated as a standard blackbody, is obtained fitting the optical spectrum. We notice that, for the cases considered in \S~\ref{Sezione_4}, we need three different IR (dust) components and one optical component. The presence of three separate populations of dust is probably unphysical, but here are used to provide a good fit to the spectra.

The emissivity of ICS \citep[see][]{PhysRev.167.1159} is computed under the assumption that the low energy background photon field is concentrated at the peak $\epsilon_{\rm peak}$ of the dust and starlight components. This approximation leads a factor $\sim 2$ of uncertainty in the predicted IC flux. We consider this uncertainty as acceptable since IC is subdominant compared to other channels.
Within this approximation the IC emissivity is given by:
\begin{align}
\begin{split}
\phantom{xxxxxxxxxx} 4 \pi \; \epsilon_{\rm IC}(E, \epsilon_{\rm peak},r) = 
				 \frac{3 c \sigma_{\rm T}}{4} \frac{U_{\rm rad}}{\epsilon_{\rm peak}^2} \; \times \\ \int^{\infty}_{p_{\min}} f_e(p,r)
				  \left[ \frac{m_e c^2}{E_e(p)} \right]^2 G\left(q,\Gamma \right) \,4 \pi p^2 dp,
\end{split}
\end{align}
where $U_{\rm rad}$ is the energy density of the thermal component, $f_e(p,r)$ is the electron distribution function (primary + secondaries), $p_{\min}$ is the momentum corresponding to the threshold energy $E_e$ such that $E_e=  E/2 \left[1+\left(1+m_e^2 c^4/(E \epsilon_{\rm peak}) \right)^{1/2} \right]$. The function $G(q,\Gamma)$ and the variables $q$ and $\Gamma$ are reported in Appendix \ref{app:timescales}. The luminosity of each thermal component "$i$" is computed as $U_{\rm rad,i}= 9 L_{i}/(16 \pi R^2 c)$, namely assuming that the spherical SBN is not opaque at those wavelengths \citep[see also][\S~1.6]{2013LNP...873.....G}.

Gamma rays with energy above threshold for pair production may be absorbed inside the SBN, and in turn lead to the production of (tertiary) electrons (and positrons). In the same way, low frequency radiation may be absorbed due to free-free absorption whose emissivity is given by:
\begin{equation}
\phantom{xxxxxxx} \epsilon_{ff}(E)= 6.8 \times 10^{-38} T^{-1/2} Z^2 n_e n_i e^{-E/kT} \bar{g}_{ff},
\end{equation}
where $\bar{g}_{ff}$ is the mean Gaunt factor \cite[]{1986rpa..book.....R,1973blho.conf..343N} in a plasma with temperature $T$, $Z$ is the electric charge of the plasma elements, namely protons and electrons (with densities $n_i=n_e$).

In order to account for absorption, the flux of radiation escaping the SBN is calculated by solving the radiative transfer equation in the whole starburst nucleus \cite[see, e.g.][]{1986rpa..book.....R}:
\begin{equation}
  \phantom{xxxxxxxxxxx}  \frac{dI(E,s)}{ds}= \epsilon(E) - I(E,s) \eta(E) ,
  \label{radiative_transfer}
\end{equation}
where $\eta$ is the absorption coefficient for photons of given energy $E$. In the high energy part of the spectrum it takes into account $\gamma \gamma$ absorption, $\eta= \eta_{\gamma \gamma}= \int \sigma_{\gamma \gamma}(E,E') n_{\rm bkg}(E') dE'$, whereas at low energies it describes free-free absorption $\eta=\eta_{ff} \approx 0.018 T^{-3/2} Z^2 n_e n_i \bar{g}_{ff} (h/E)^2$. The spatial coordinate $s$ runs through the SBN at a given distance from the center. 

The intensity $I(E)$ for each line of sight across the SBN is calculated by solving numerically Eq.~\ref{radiative_transfer} then, summing up over all line of sight we get the total luminosity of the SBN.


Although redshift effects from nearby SBNi are negligible, absorption of gamma rays at very high energies due to pair production off the diffuse background light remains important at it is accounted for following the approach of \citet{2017A&A...603A..34F} (see Appendix \ref{app:EBL} for the detailed description).


\section{CR Diffusion and calorimetry}
\label{Sezione_3}

The modelling of the non thermal activity of SBGs relies upon the assessment of the assumption of calorimetry, that is often adopted with no much discussion in most literature on the topic. In this section we address the issue of whether CRs lose most of their energy inside SBNi or not in terms of CR transport, and we discuss the observational evidence in terms of emission of non thermal radiation. In order to reach this goal, we compute the spectra of protons, (primary, secondary and tertiary) electrons and the radiation emitted by them in the three diffusion models discussed earlier. 

Model A is our benchmark transport model: it predicts that protons lose an appreciable fraction of their energy inside the SBN, although the time scale for escape is comparable with that of the wind advection. The smallness of the diffusion coefficient for this model causes the advection to be the main channel of escape of CR protons from the nucleus, for energies as high as $\sim 10$ PeV. The time scale of energy losses of protons, dominated by pion production, becomes shorter than the advection time above $\sim 10$ TeV, because of the weak energy dependence of the cross section for this process. 

All electrons (primary, secondary from pp collisions and tertiary) lose their energy inside the SBN, hence the assumption of calorimetry is certainly justified for the electrons. 



\begin{figure}
\centering
\includegraphics[width=0.45\textwidth]{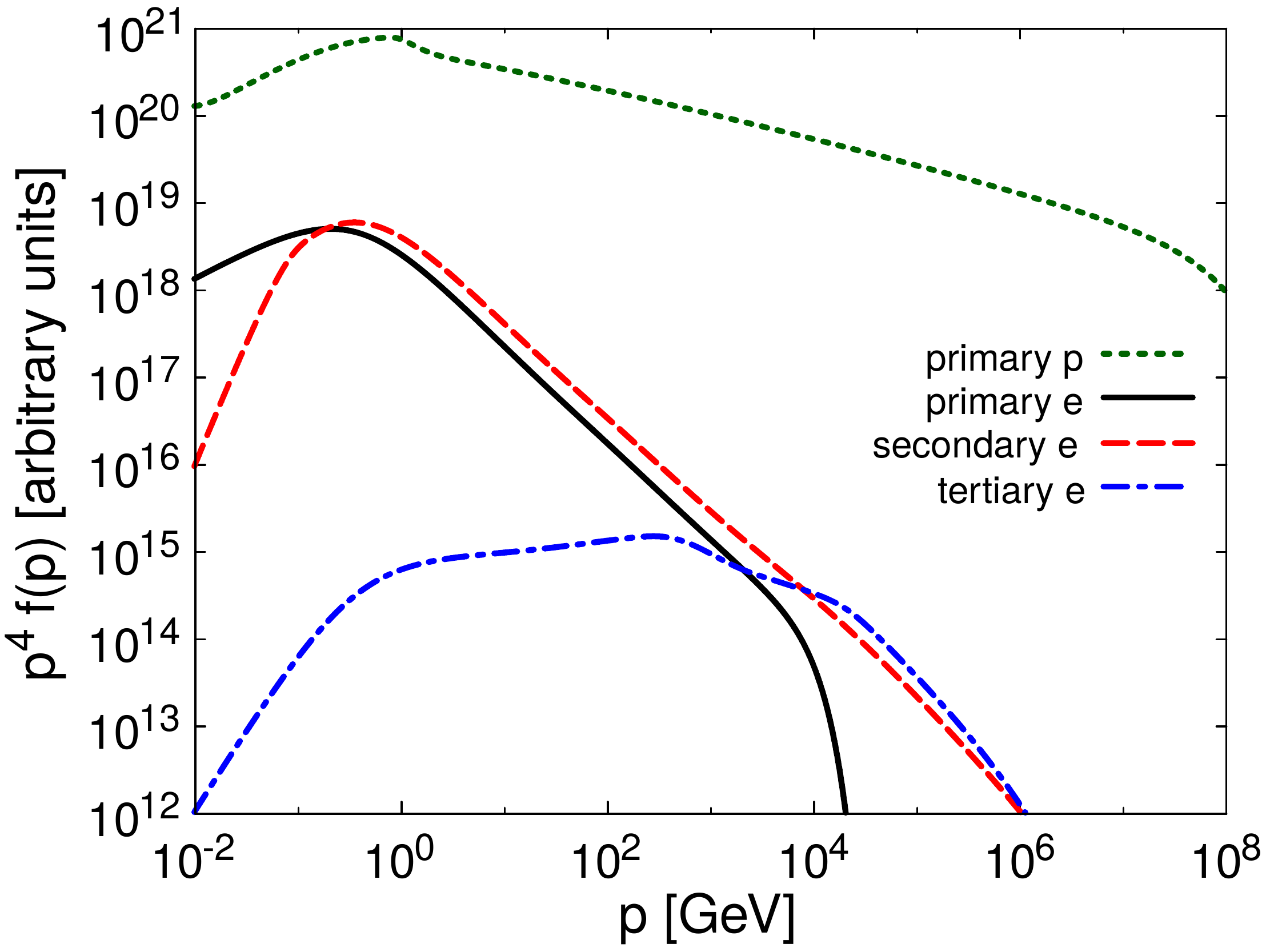}\quad\includegraphics[width=0.45\textwidth]{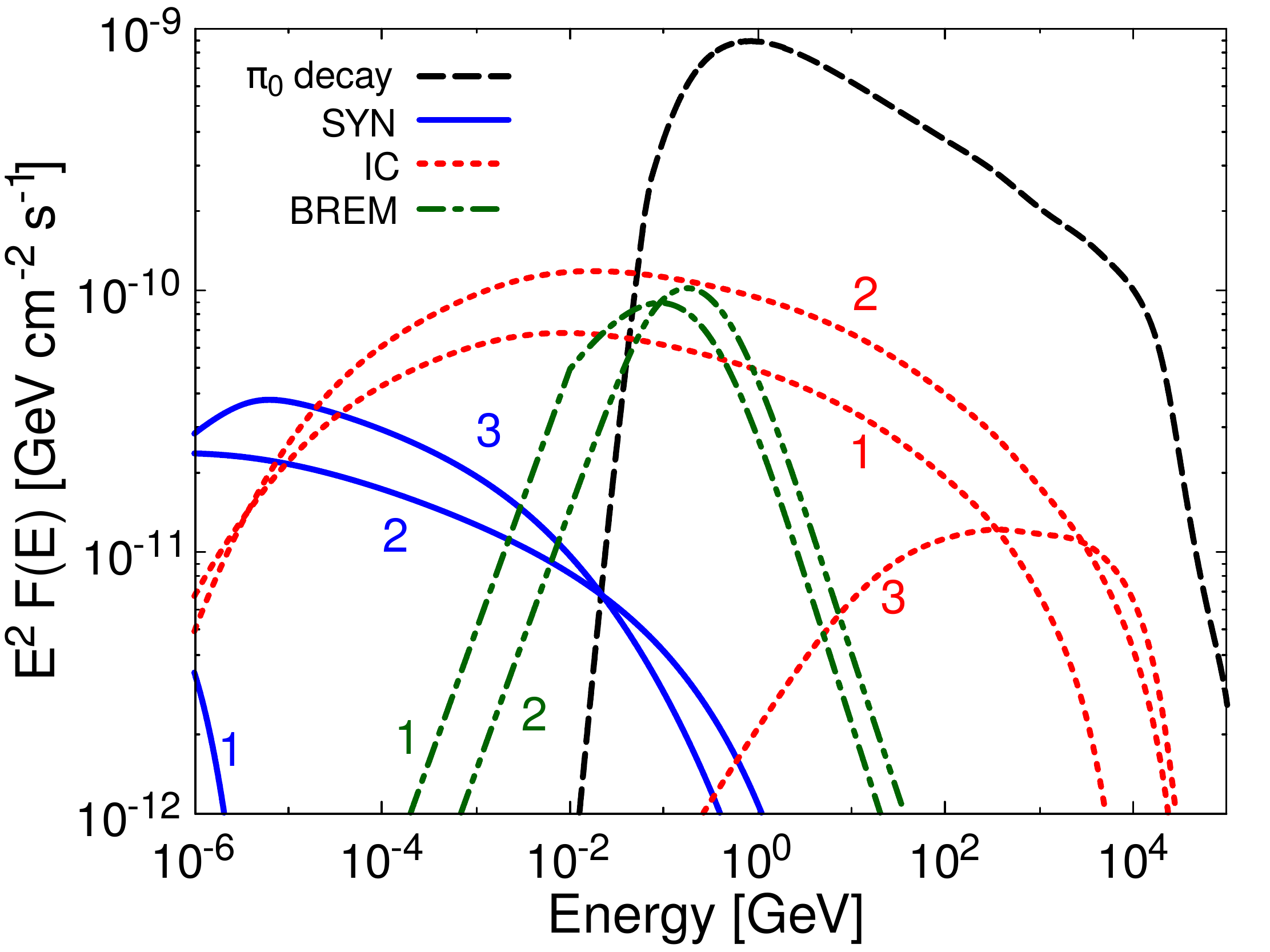}
\caption{\label{fig:Kolmogorov} Particle and photon spectra in Model A. The upper panel shows primary protons (green dashed), primary electrons (black thick line), secondary electrons (red dashed line) and tertiary electrons (blue dot-dashed line). The lower panel shows the high energy spectral components of $\pi_0$ decay (black dashed line), inverse Compton (red dotted line), synchrotron (blue thick line) and bremsstrahlung (green dot-dashed line). The relative contributions of the different electron populations are separated in primaries ($1$), secondaries ($2$) and tertiaries ($3$).}
\end{figure}

The equilibrium spectra of protons and electrons for Model A are shown in Figure \ref{fig:Kolmogorov} (top panel), where we adopted the reference values of parameters as listed in Table \ref{tab:parameters_table}. The strong role of energy losses makes the spectrum of protons reflect the injection spectrum, with a small correction due to the energy dependence of losses. For electrons, energy losses are always faster than both advection and diffusion, hence their spectrum is steeper than the injection spectrum by approximately one power of energy, since the main channel of losses are represented by synchrotron emission in the intense magnetic field of the SBN and by IC off the IR photons. Small wiggles are present in the high energy spectrum of secondary electrons (not clearly visible in the Figure due to the large vertical scale) reflecting the fact that the cross section for ICS off photon backgrounds enters the Klein-Nishina regime when $E_{\gamma}\epsilon_{ph}\sim m_{e}^{2} c^4$. At energies $\lesssim$ GeV, the spectrum of primary electrons is dominated by ionization losses, while for secondary electrons the low energy part of the spectrum falls fast because of the threshold for pion production in pp collisions. Tertiary electrons, produced by pair production of high energy gamma rays in the SBN, start at energies $\sim$ TeV, where absorption off the NIR background becomes important. A second peak is present at energies $\sim 20$ TeV due to the peak in the FIR. On the contrary the contribution due to optical photons is almost negligible. The spectrum of tertiary electrons at energies lower than their minimum injection energy is due to synchrotron and ICS ageing of tertiary electrons injected at higher energies.  

The bottom panel of Figure \ref{fig:Kolmogorov} shows the photon spectra from a SBN with the values of the parameters listed in Table \ref{tab:parameters_table}. The number labels refer to the contribution of primary (1), secondary (2) and tertiary (3) electrons. Most gamma rays with energy $\gtrsim 100$ MeV are due to production and decay of neutral pions. The cutoff in the spectrum of gamma rays at energies $\sim 10$ TeV is due to absorption of gamma rays inside the SBN. For larger distances of the galaxies from the Earth the absorption on the extragalactic background light is also expected to become important. We will discuss this point further when dealing with individual sources. 

It is interesting to notice that while the synchrotron emission of primary electrons quickly becomes unimportant at high photon energy, the synchrotron emission of secondary and tertiary electrons is dominant in the hard X-ray band. Hence the detection of such hard X-ray emission may be considered as a rather unique signature of strong CR interactions inside the SBN and corresponding copious production of secondary electrons and even efficient gamma ray absorption (tertiary electrons). 

In the soft gamma ray band, most emission is due to a combination of ICS and bremsstrahlung of primary and secondary electrons and to synchrotron emission of secondary and tertiary electrons. 

\begin{figure}
\centering
\includegraphics[width=0.45\textwidth]{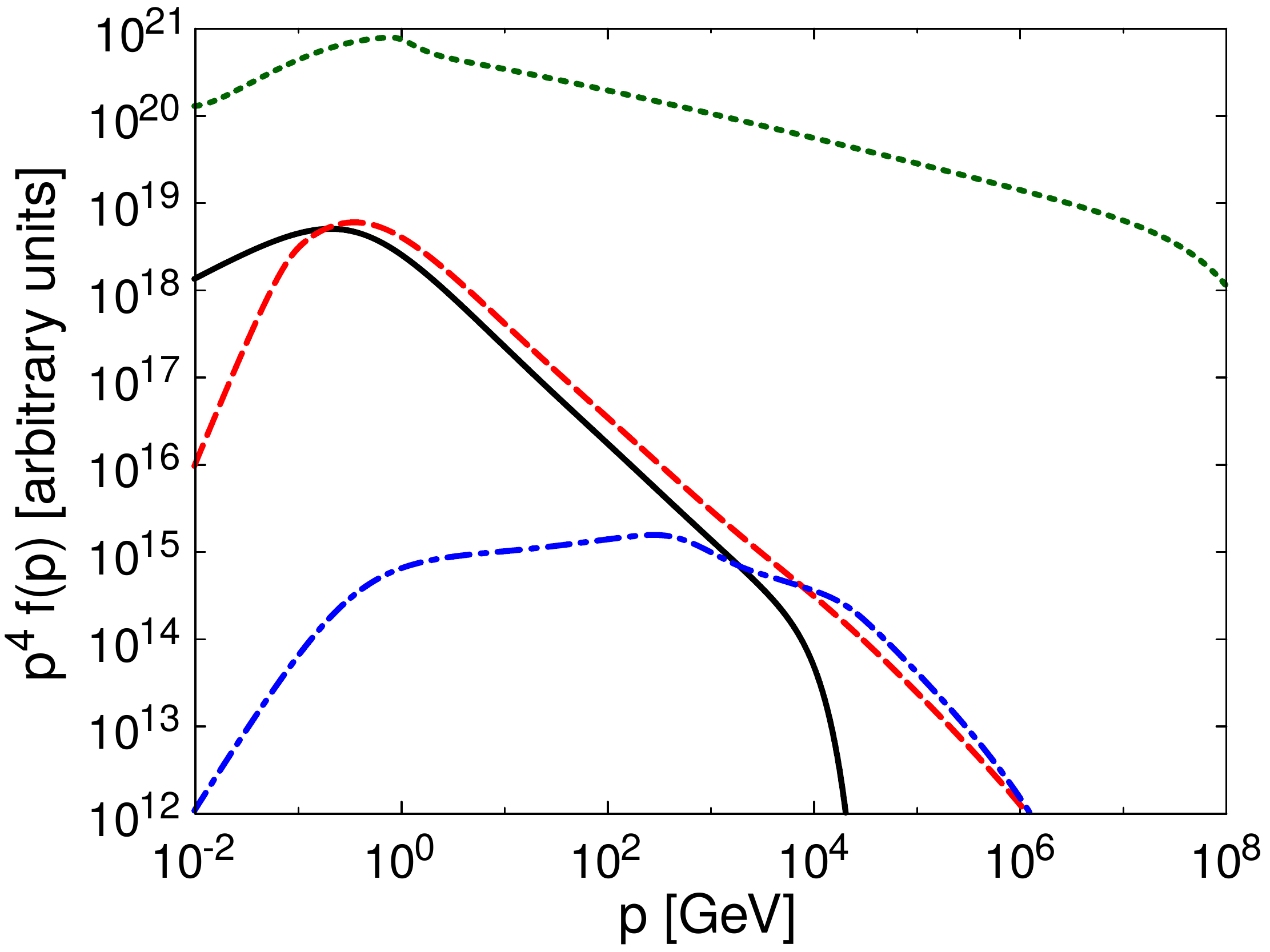}\quad\includegraphics[width=0.45\textwidth]{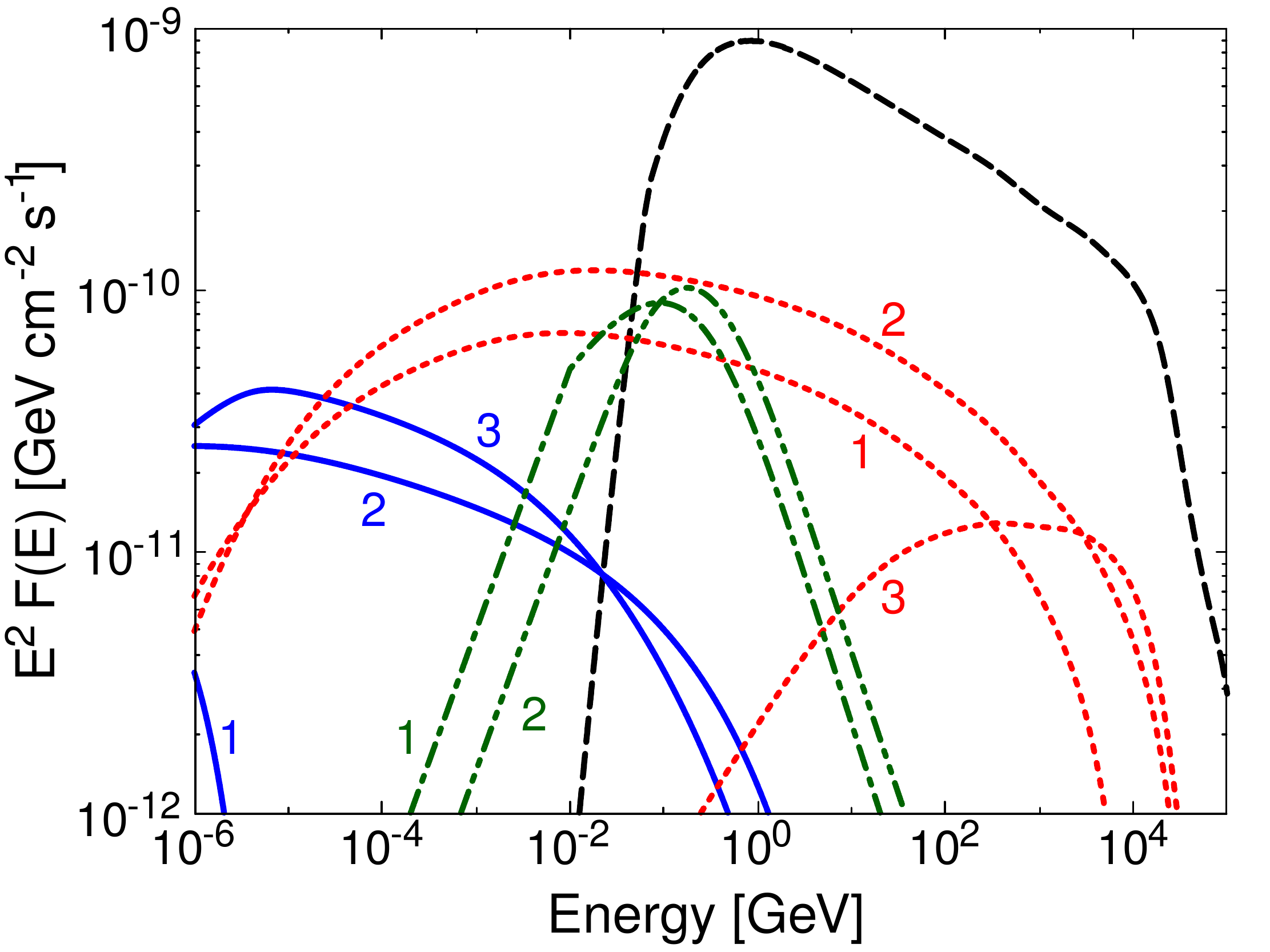}
\caption{\label{fig:Bohm}  Particle and photon spectra in Model B. The line style is the same as in Figure \ref{fig:Kolmogorov}.}
\end{figure} 

\begin{figure}
\centering
\includegraphics[width=0.45\textwidth]{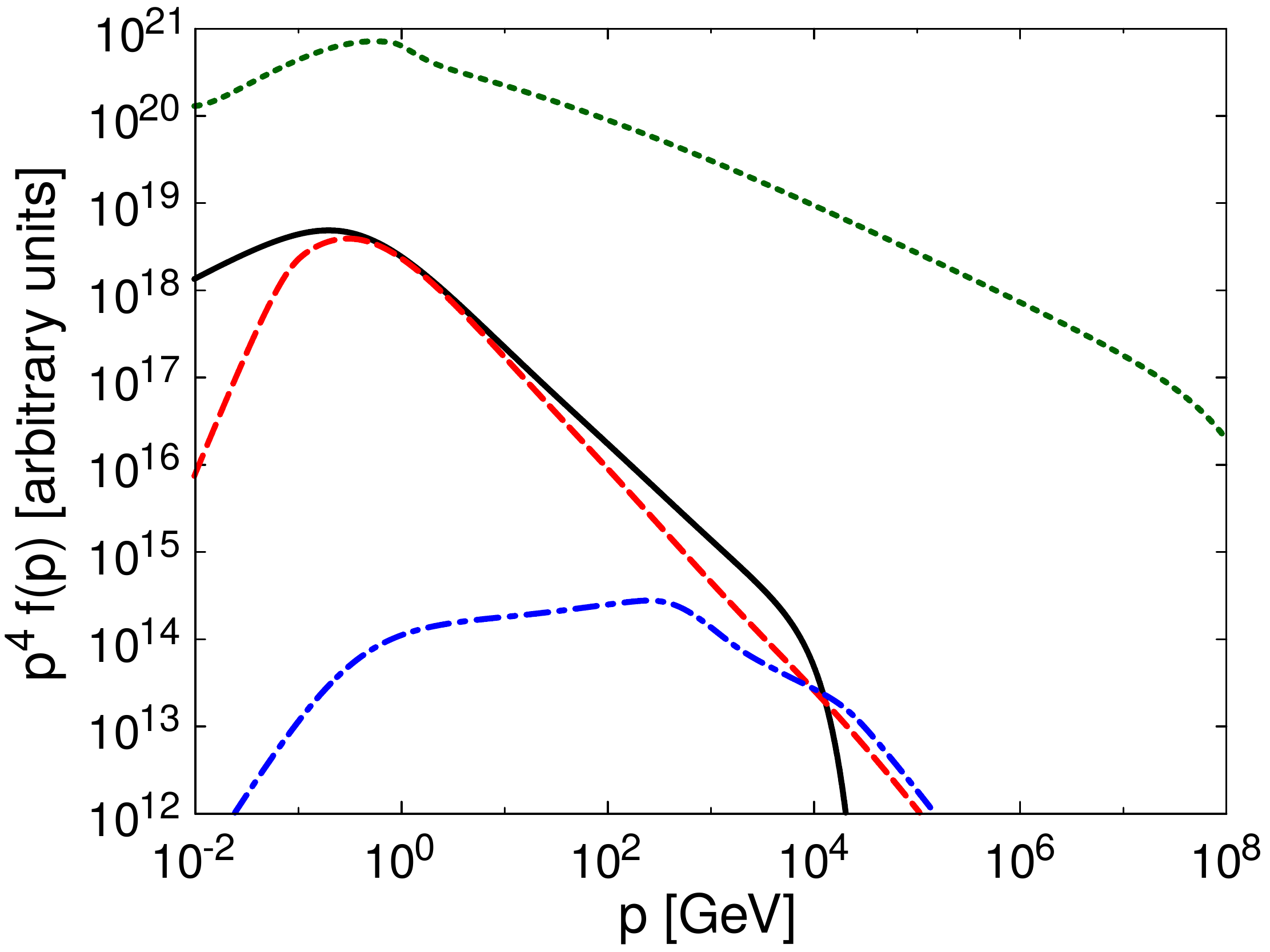}\quad\includegraphics[width=0.45\textwidth]{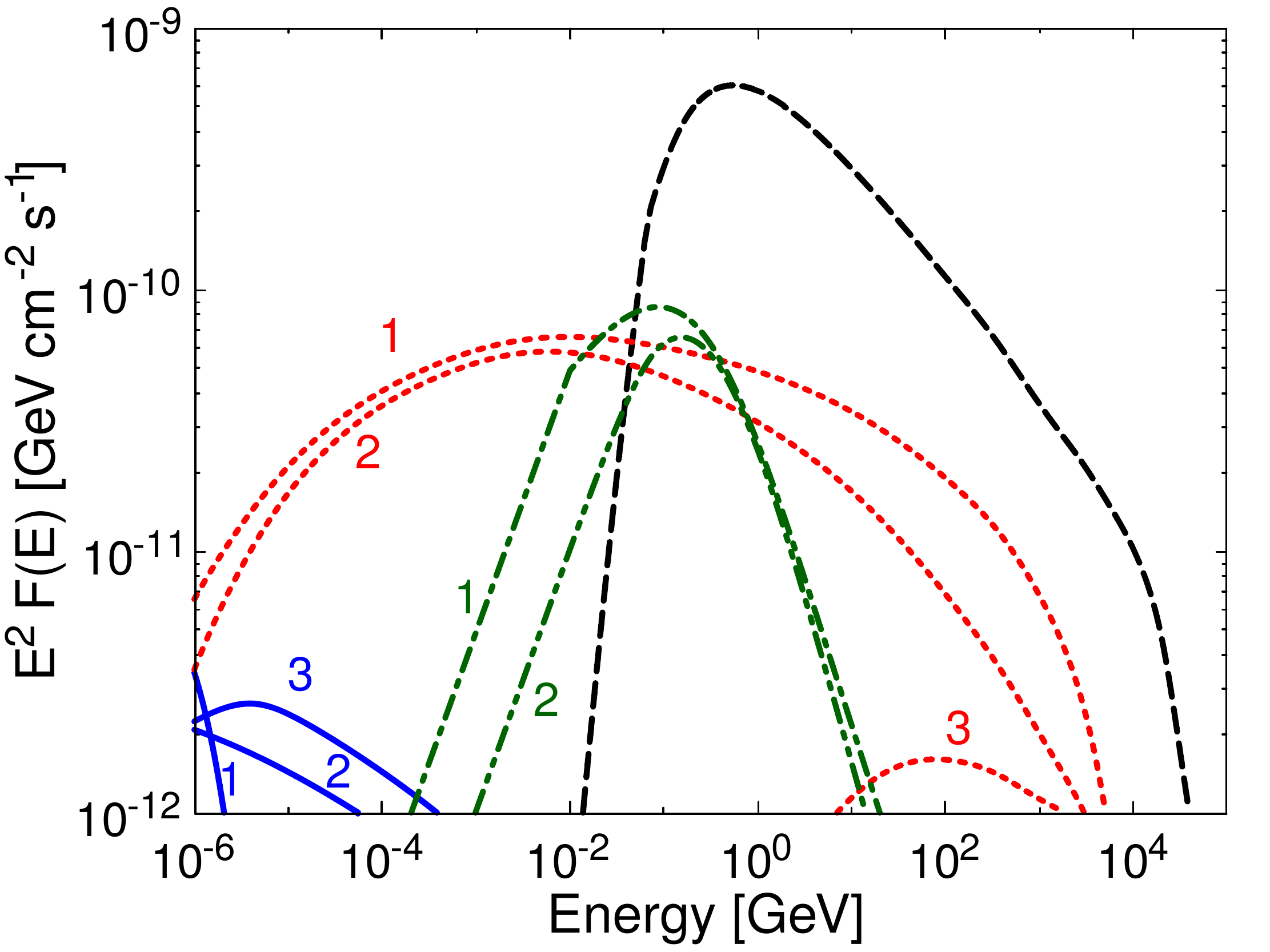}
\caption{\label{fig:MW-like} Particle and photon spectra in Model C. The line style is the same as in Figure \ref{fig:Kolmogorov}.}
\end{figure}

In Figure \ref{fig:Bohm} we show the particle (upper panel) and the photon spectrum (lower panel) in the context of Model B, where Bohm diffusion was assumed. Although the time scale for diffusion in Model B is typically much shorter than for Model A, not much difference is observed in the predicted spectra, as a result of the fact that in both models the transport of CRs is mostly dominated by advection and energy losses. Electrons are well confined inside the SBN and lose all their energy inside the nucleus. These two conditions imply that calorimetry is a good approximation for both Model A and B, hence much of what has been said for Model A also applies to Model B. 

Model C is qualitatively different from previous diffusion models, in that the larger diffusion coefficient determines a transition from advection to diffusion dominated transport at $E\sim 1$ GeV for protons, while electrons remain loss dominated. The corresponding results are shown in Figure~\ref{fig:MW-like}. The spectrum of CR protons is steeper than injection by an amount determined by the energy dependence of the diffusion coefficient (1/3) and, as a consequence, the injection spectrum of secondary electrons is correspondingly steeper as $ E^{-(2.25+1/3)}$. Moreover, the shorter diffusion time leads to a smaller density of secondary electron when compared with the results of Models A and B, so that the electron spectrum is now dominated by primary electrons. 
The main imprints on the spectrum of photons (lower panel) are the steeper spectrum of gamma rays from $\pi^{0}$ decays and the fact that the synchrotron emission in the hard X-ray band is sizeably smaller than for Models A and B, as a result of the lack of calorimetry for CR protons. 

The different emission in the hard X-ray band between Models A and B on one hand and Model C on the other illustrate well the potential importance of the detection of hard X-rays from SBNi, in that such photons carry information about the calorimetric properties of the SBN. 
Although hard X-rays from the cores of SBGs have been observed \cite[]{2007ApJ...658..258S,2017ApJ...841...44P,2014ApJ...797...79W}, an important contribution to such diffuse emission is typically attributed to unresolved X-ray binaries (XBs), SNRs, $O$ or early-$B$ spectral type stars, diffuse thermal plasma and a possible AGN activity \citep[for a detailed discussion of these components see][]{2002A&A...382..843P}. CR electrons are also expected to contribute to the diffuse hard X-ray emission mainly through ICS on the IR background  \citep[see ][]{2002A&A...382..843P}. The possibility that a contribution to the diffuse hard X-ray flux could come from synchrotron emission of CR electrons was first suggested by \citet{0004-637X-762-1-29}. Nevertheless, in their model the X-ray emission is dominated by IC  from primary electrons and is roughly 10 time smaller than our prediction in the same energy band, which is, instead, dominated by synchrotron emission from secondary and tertiary electrons. In their case the contribution from secondary electrons is much smaller due to a faster diffusion of protons.
Indeed, as we discussed earlier, the contribution of secondary and tertiary electrons to the diffuse hard X-ray emission reflects the effectiveness of the confinement of CRs inside SBNi, which in turn can be expressed in terms of luminosity in some selected bands. 

\begin{table}
\centering
\begin{tabular}{|c|c|c|c|}
\hline
  &  Model A  & Model B & Model C \\ \hline \hline
 $L_{\gamma}$	& $162 $ & $163 $ & $94 $   \\ \hline
 $L_{\rm IR}$  & $1.65 \times 10^{6}$  & $1.65 \times 10^{6}$  & $1.65 \times 10^{6}$   \\ \hline
 $L_{X}$ 	& $13.6 \; [7.1,6.5] $ & $ 14.3 \; [7.8,6.5] $ & $ 5.6 \; [0.5,5.1] $   \\ \hline
 $L_{X_1}$ & $4.8 \; [3.4,1.4]$ & $ 5.1 \; [3.7,1.4]$ & $ 1.5 \; [0.3,1.2]$   \\ \hline
 $L_{X_2}$ & $5.4 \; [3.0,2.4]$ & $ 5.7 \; [3.2,2.5]$ & $ 2.1 \; [0.2,1.9]$   \\ \hline
 $L_{X_3}$ & $5.3 \; [2.0,3.3]$ & $ 5.5 \; [2.2,3.3]$ & $ 2.6 \; [0.1,2.5]$   \\ \hline
\end{tabular}
\caption{\label{tab:outcomes_diffusion_table} Luminosity (expressed in units of $10^{38} erg/s$) in three selected energy bands in Models A, B and C. $L_{\gamma}$ is the gamma-ray luminosity computed in the energy range $0.1-10^2 GeV$, whereas $L_{\rm IR}$ is computed in the far infrared ($8 \; \mu m<\lambda<10^3 \mu m$). $L_X$ is computed in the X-ray channel $1-10^2$ keV, whereas $L_{X_{1}}$, $L_{X_2}$ and $L_{X_3}$ are computed in the sub-bands $1-8$ keV, $4-25$ keV and $25-100$ keV, respectively. The square brackets show separately the contribution of SYN and IC to the total luminosity (value out of the parentheses).}
\end{table} 

In Table~\ref{tab:outcomes_diffusion_table} we show the luminosity in gamma-rays ($0.1-10^2$ GeV), X-rays ($1-10^2$ keV ) and IR radiation ($8-10^3 \mu \rm m$). Models A and B basically return the same result. On the other hand, Model C shows a clear reduction in the X-ray and gamma-ray luminosity by about a factor $\sim 2\div 3$, while the IR luminosity remains unchanged since the thermal contribution dominates upon synchrotron by $\sim 5$ orders of magnitudes.
For completeness, in the same Table, we also report the X-ray luminosities in three bands, $1-8$ keV (typical of Chandra), at $4-25$ keV (typical of NuStar) and $25-10^2$ keV. 
Clearly, the synchrotron emission of secondary and tertiary electrons can contribute (together with the XRBs component) to provide a natural explanation of the hard X-ray extra-component in the band $0.5-10$ keV discussed in \citet{2002A&A...382..843P}.

\section{Application to known SBGs}
\label{Sezione_4}


In this section we specialize our calculation to the case of the three nearby SBGs, namely NGC253, M82 (respectively with \citealp[$D_L \approx 3.8$ and $D_L \approx 3.9$ as found by][]{2005MNRAS.361..330R,1999ApJ...526..599S}) and Arp220 \citep[located at $D_L \approx 77 \; Mpc$, as inferred by][]{1538-4357-492-2-L107}. The latter belongs to the ULIRG class, characterized by very prominent IR luminosity, higher ISM density and magnetic field energy density and more intense star formation activity \cite[]{2015ApJ...800...70S}. Arp220 shows a rate of SN explosions which is more than one order of magnitude higher than typical SBGs \cite[]{2003A&A...401..519M,2006ApJ...647..185L}. 


For the modelling of the emission from these SBGs we start by fitting the thermal emission, in the $\sim 0.1$ meV - few eV range, assuming that the observed emission in this band is dominated by the SBN and then we tune the other parameters to fit the multiwavelength spectra, from radio to gamma rays. The parameters' values used for each source are listed in Table \ref{tab:input_fits}.


\begin{table}
\centering
\begin{tabular}{|c|c|c|c|}
\hline
 Parameters  &  NGC253 & M82  & Arp220  \\ \hline	\hline
 $D_L$ (Mpc) [z] & $3.8$ $[8.8 \; 10^{-4}]$ & $3.9$ $[9 \; 10^{-4}]$  & $77.0$ $[1.76 \; 10^{-2}]$ \\ \hline
 $\mathcal{R}_{\rm SN}$ (yr$^{-1}$) & $ 0.027$ & $0.05$ & $ 2.25$ \\ \hline
 $R$ (pc) & $150$ & $220$ & $250$  \\ \hline
 $\alpha$  & $4.3$ & $4.25$ & $ 4.45$ \\ \hline
 $B$ ($\mu$G) & $ 170$ & $ 210$ & $ 500$  \\ \hline
 $M_{\rm mol}$ $(10^8 M_{\odot})$ & $0.88$ & $1.94$ & $ 57 $ \\ \hline 
 $n_{\rm ISM}$ (cm$^{-3}$) & $ 250$ & $175$ & $ 3500$ \\ \hline
 $n_{\rm ion}$ (cm$^{-3}$) & $ 30$ & $22.75$ & $ 87.5$  \\ \hline
 $v_{\rm wind}$  (km/s) & $300$ & $600$ & $500$ \\ \hline
 $T_{\rm plasma}$ (K) & $8000$ & $7000$ & $3000$ \\ \hline 
 $U^{\rm FIR}_{\rm eV/cm^3}$ [$
 \frac{\rm kT}{\rm meV}$] & $ 1958$ $[3.5]$ & $ 910$ $[3.0]$ & $ 31321$ $[3.5]$ \\ \hline
 $U^{\rm MIR}_{\rm eV/cm^3}$ [$
 \frac{\rm kT}{\rm meV}$] & $ 587$ $[8.75]$ & $ 637$ $[7.5]$ & $ 9396$ $[7.0]$ \\ \hline
 $U^{\rm NIR}_{\rm eV/cm^3}$ [$
 \frac{\rm kT}{\rm meV}$] & $ 587$ $[29.75]$ & $ 455$ $[24.0]$ & $ 125$ $[29.75]$ \\ \hline
 $U^{\rm OPT}_{\rm eV/cm^3}$ [$
 \frac{\rm kT}{\rm meV}$] & $ 2936$ $[332.5]$ & $ 546$ $[330.0]$ & $ 1566$ $[350.0]$ \\ \hline
\end{tabular}
\caption{\label{tab:input_fits} Input parameters for the galaxies examined in \S~\ref{Sezione_4}.}
\end{table}

We check {\it a posteriori} that the best fit values of the parameters are in agreement with values reported in the literature. In particular, the inferred radius of the SBN and the ISM conditions appear to compare well with the values presented by \citet{0004-637X-735-1-19} and \citet{2008ApJ...689L.109H} for NGC253, \citet{2003ApJ...599..193F} and \citet{2001ApJ...552..544F} for M82. 
For Arp220, we adopt a simplified spherical geometry embedding the two galactic nuclei that are observed. In fact we have adopted parameters that are a reasonable average between the highly compact SBNi and their surrounding environment (\citealp[detailed observations of Arp220 and its ISM condition are discussed in][]{2015ApJ...800...70S,1999ApJ...514...68S}). Parameters like the average magnetic field and the advection speed have been taken consistently with typical values expected from SBNi (\citealp[see for instance][]{2006ApJ...645..186T,2017arXiv170109062H} respectively).



For all three sources analyzed in this section, radio data in the frequency range $1-10$ GHz are taken from \cite[]{Radio_sources}, whereas data at higher energies, namely from $\sim 0.1$ meV to $\sim 10$ eV, have been retrieved from the NED\footnote{The NASA/IPAC Extragalactic Database (NED) is operated by the Jet Propulsion Laboratory, California Institute of Technology, under contract with the National Aeronautics and Space Administration.} catalog (in particular we use the SED-builder online tool https://tools.ssdc.asi.it/). In all cases we use three DBBs for the dust contribution, a normal BB for the stellar component and a free-free contribution from the thermal plasma. The parameters of these low energy components are listed in the last five rows of Table \ref{tab:input_fits} while Table~\ref{tab:results}  summarizes the main outcomes of our modelling for all the three SBNi. Below we briefly describe our findings related to the three chosen SBG and we draw some general conclusions. 


\paragraph*{NGC253:} The nuclear region of NGC253 is very compact and luminous at optical wavelength. This causes a non-negligible $\gamma \gamma$ absorption at energies of few hundred GeV, which in turn determines a softening of the gamma-ray spectrum already below $\sim 1$ TeV. The spectrum above $100$ MeV is totally dominated by the $\pi_0$ component. Below $\sim 100$ MeV the dominant emission mechanism is IC (mainly from secondary electrons). Only at keV energies the IC emission becomes comparable with the SYN components from secondary and tertiary electrons. Relativistic bremsstrahlung is always subdominant but provides a non negligible contribution to the total gamma ray emission in the range $10 \div 100$ MeV. 

\begin{figure}
\centering
\includegraphics[width=0.45\textwidth]{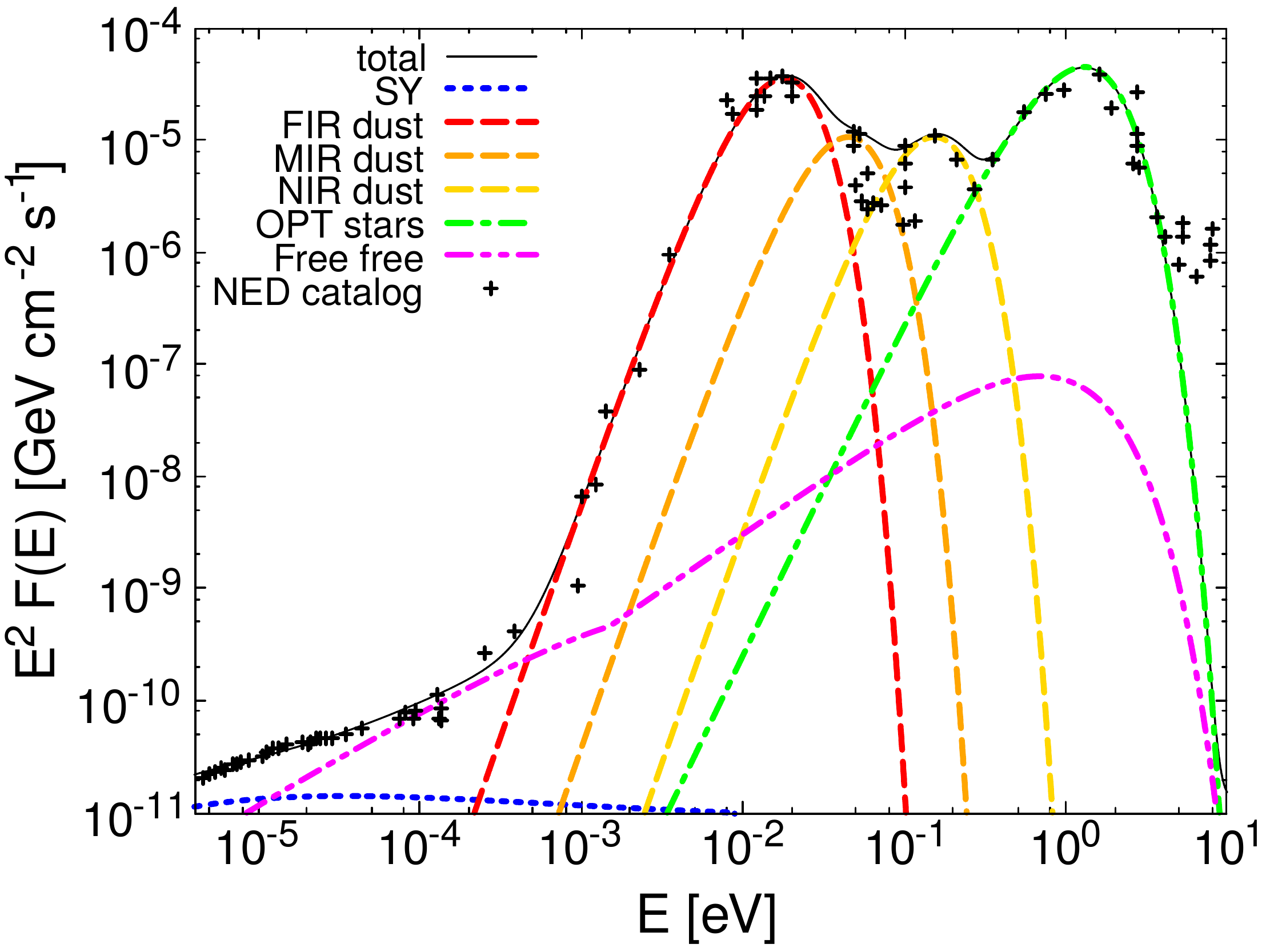}\quad\includegraphics[width=0.45\textwidth]{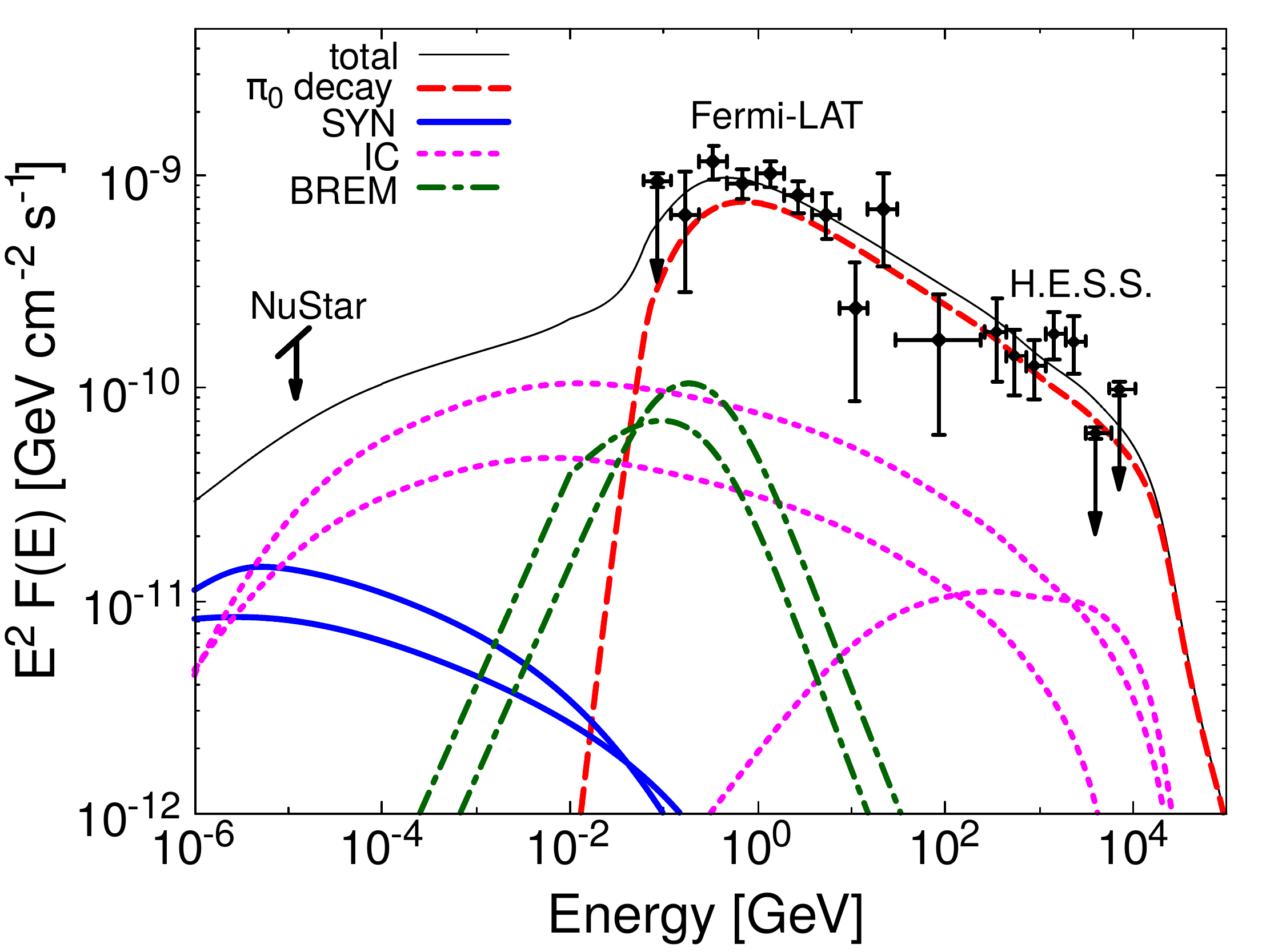}
\caption{\label{fig:NGC253} Multiwavelength spectrum of NGC253. Upper panel shows the low energy spectrum with relative components:: thermal dust DBBs (red, orange and yellow dashed), optical star BB (green dot-dashed), thermal free-free (magenta dot-dot-dashed) and SYN (blue dotted). Lower panel shows the high energy spectral components: $\pi_0$ (red dashed), IC (magenta dotted), BREM (green dot-dashed) and SYN (blue dashed). Together with the photons we show the single flavor neutrino flux (thin gold dashed). The data (black points) are observed by Fermi-LAT and HESS and presented in \citet{Abdalla:2018nlz} for the HE and VHE domain, whereas the hard X-ray upper limit is taken from \citet{2014ApJ...797...79W}.}
\end{figure}

The multifrequency spectrum of NGC253 is shown in Figure~\ref{fig:NGC253}. The top panel illustrates the good agreement between the results of our modelling of the low energy emission and observations. The bottom panel is more interesting in that it shows the gamma ray emission coming from both the decays of neutral pions and from interactions of electrons and gamma rays with magnetic fields and low energy photon background inside the SBN. 

Gamma-ray data collected by Fermi-LAT and HESS \citep[see][]{Abdalla:2018nlz} are well reproduced. Of particular interest is the shape of the spectrum below $\sim 1$ GeV where data show a strong hint of the pion bump, a clear signature of the hadronic origin of gamma-rays. The computed hard X-ray flux, contributed by both synchrotron and IC, is at the level of $E^2F(E) \approx 10^{-10} \rm GeV \, cm^2 \, s^{-1}$ at $10$ keV, appreciably larger than previous estimates \citep[e.g.][]{0004-637X-762-1-29}, but consistent with detailed observations of the nuclear region of NGC253 performed by NuStar \cite[]{2014ApJ...797...79W}. In this case, our larger flux with respect to \cite{0004-637X-762-1-29} is mainly due to the IC emission from secondary electrons copiously produced because the larger confinement time of CRs, whereas SY dominates only below $\sim 5$ keV.

\paragraph*{M82:} The multiwavelength spectrum of M82 is very similar to that of NGC253, but requires a slight harder injection (see Table \ref{tab:input_fits}) to explain the harder observed gamma-ray spectrum. In this way, the gamma-ray observations from Fermi-LAT and Veritas (\citealp[see for istance][]{2012ApJ...755..164A,2009Natur.462..770V}) are again well reproduced. The absorption of VHE gamma-rays is almost negligible below a few $TeV$ because the optical background is almost a factor $5$ lower at the peak with respect to the case of NGC253. 

\begin{figure}
\centering
\includegraphics[width=0.45\textwidth]{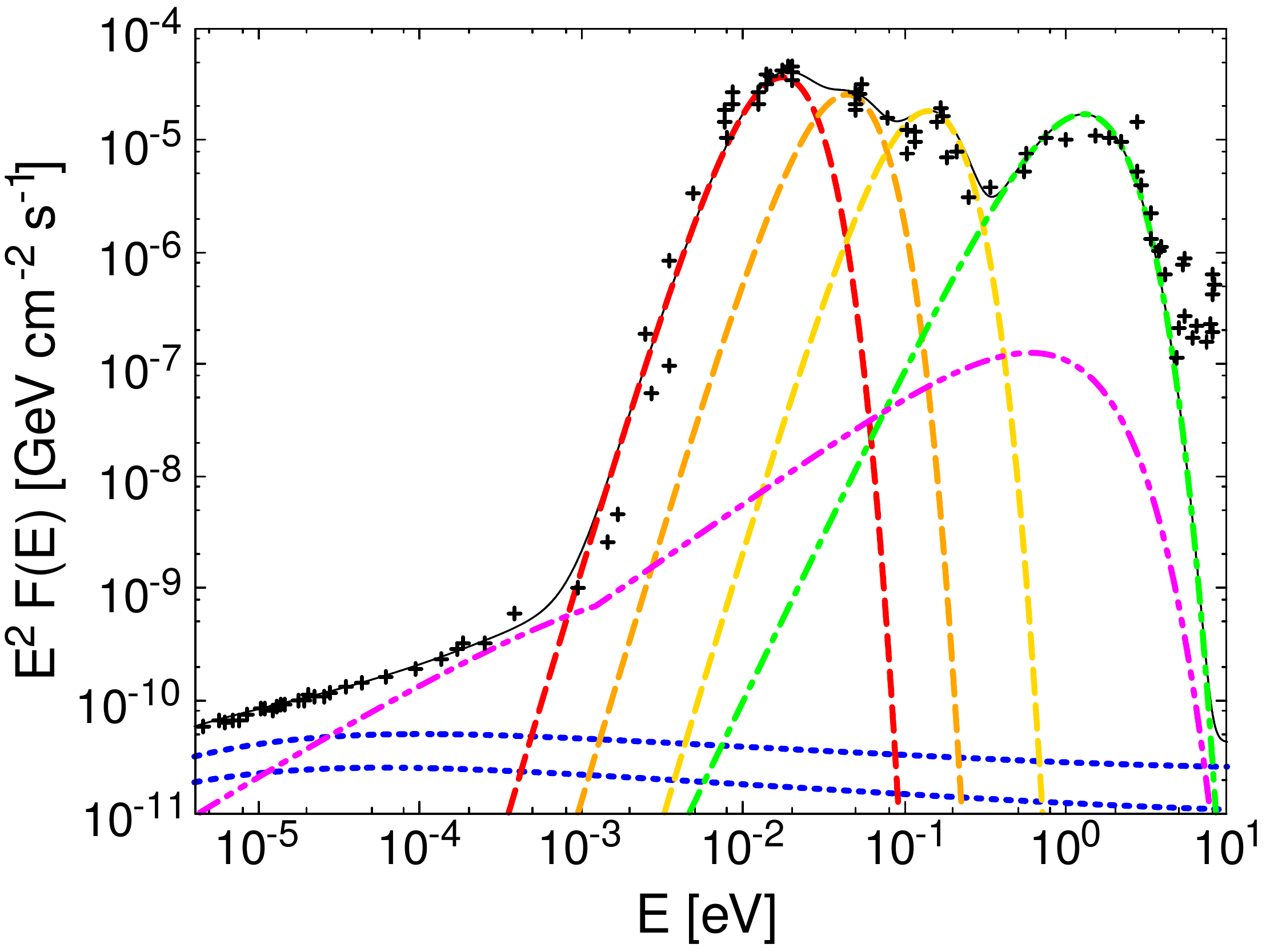}\quad\includegraphics[width=0.45\textwidth]{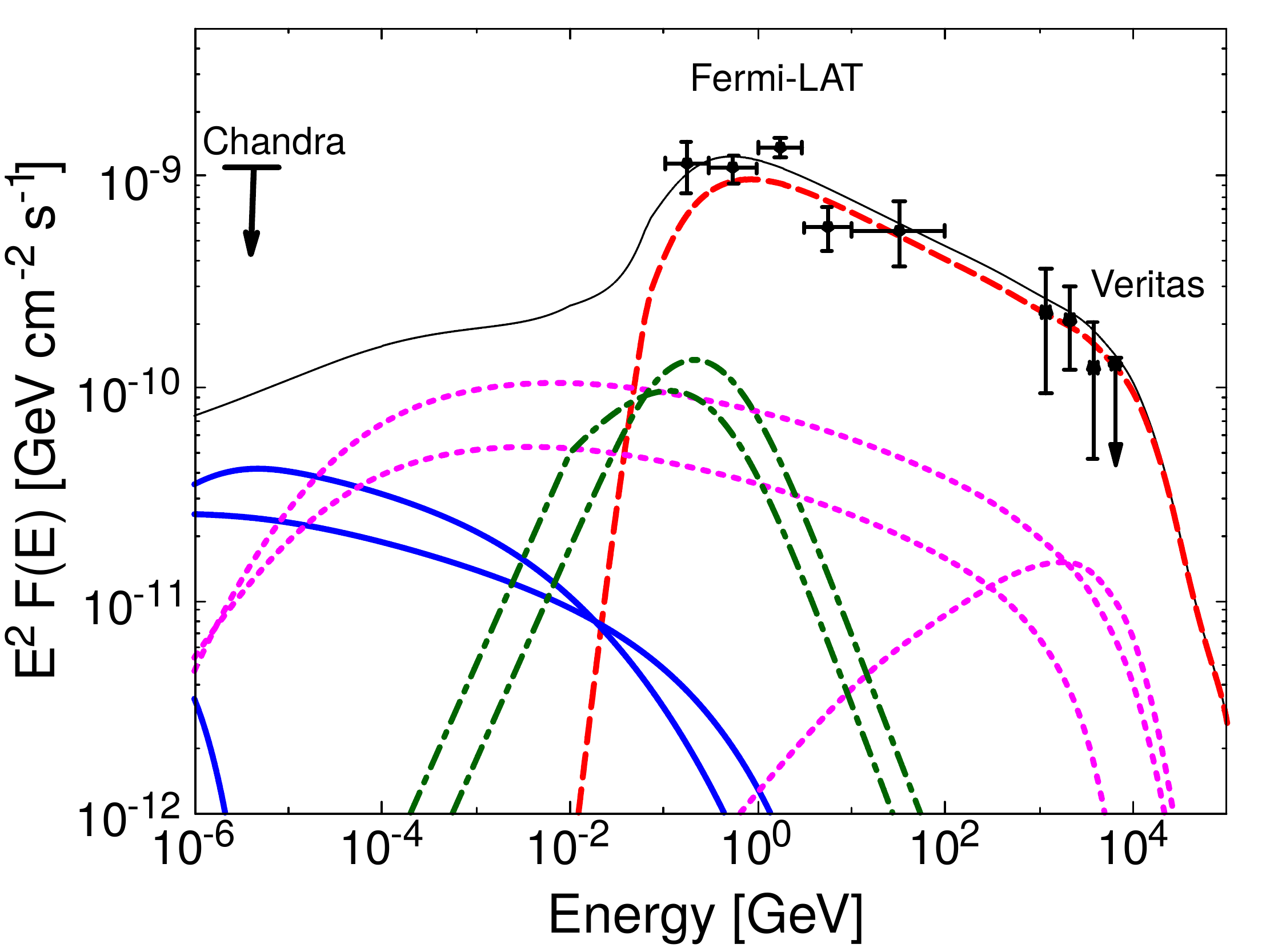}
\caption{\label{fig:M82} Multiwavelength spectrum of M82. The line style is the same of figure \ref{fig:NGC253}. The HE gamma-ray observation are taken from the Fermi-LAT observation discussed in \citet{3FGL}, whereas VHE data come from Veritas and are published in \citet{2009Natur.462..770V}. The X-ray point is a Chandra \citep[see][]{2007ApJ...658..258S} we have taken as upper limit because of possible contamination from undetected point-like sources (e.g. XRBs) and thermal plasma.}
\end{figure}

The computed diffuse hard X-ray flux is again very high ($E^2F(E)\approx 10^{-10} \rm GeV \, cm^{-2} \, s^{-1}$ at a few keV) and, different from NGC253, it is dominated by synchrotron emission of secondary and tertiary electrons up to $\sim 20$ keV. Although no measurement of the truly diffuse hard X-ray flux from the nuclear region of M82 is available at present, recent observations carried out using Chandra \citep[see][]{2007ApJ...658..258S}, XMM-Newton \citep[see ][]{2008MNRAS.386.1464R} and more recently NuStar \citep[][]{Bachetti:2014qsa} suggest that our computed hard X-ray diffuse flux is $\approx 5 \% \div 10 \%$ of the total observed flux in the energy band $3-8$ keV, hence we interpret the X-ray point in Figure \ref{fig:M82} as an upper limit to the diffuse emission, since point-like sources could contaminate such measurement. 

\paragraph*{Arp220:} 
Our simple assumptions on the geometric properties of the SBN are particularly restrictive when applied to a source such as Arp220, with its complex morphology (two nuclei and possibly a low activity AGN). In this sense, it is noteworthly that, despite such limitations, a reasonable fit to the multifrequency emission can be obtained for this sources, using the input parameters listed in the last column of Table \ref{tab:input_fits}. In particular, we have found that our best fit value for the magnetic field ($\sim 500~\mu$G) is about a factor 2 lower than the typical $\sim$mG field assumed in literature for the two SBNi of Arp220 \citep[see for istance][]{Thompson2006-magneticfield,Barcos-munoz_Arp220_B,McBride_Arp220_B,Yoast-Hull_Arp220}. Our value for the magnetic field is not in tension with previous estimates because it represents an average between the magnetic field inside the two nuclei and the one in the surrounding region, estimated to be $\sim 10^2$ $\mu$G  \citep[for similar discussions see also][]{2004ApJ...617..966T,Varenius_Arp220}.


\begin{figure}
\centering
\includegraphics[width=0.45\textwidth]{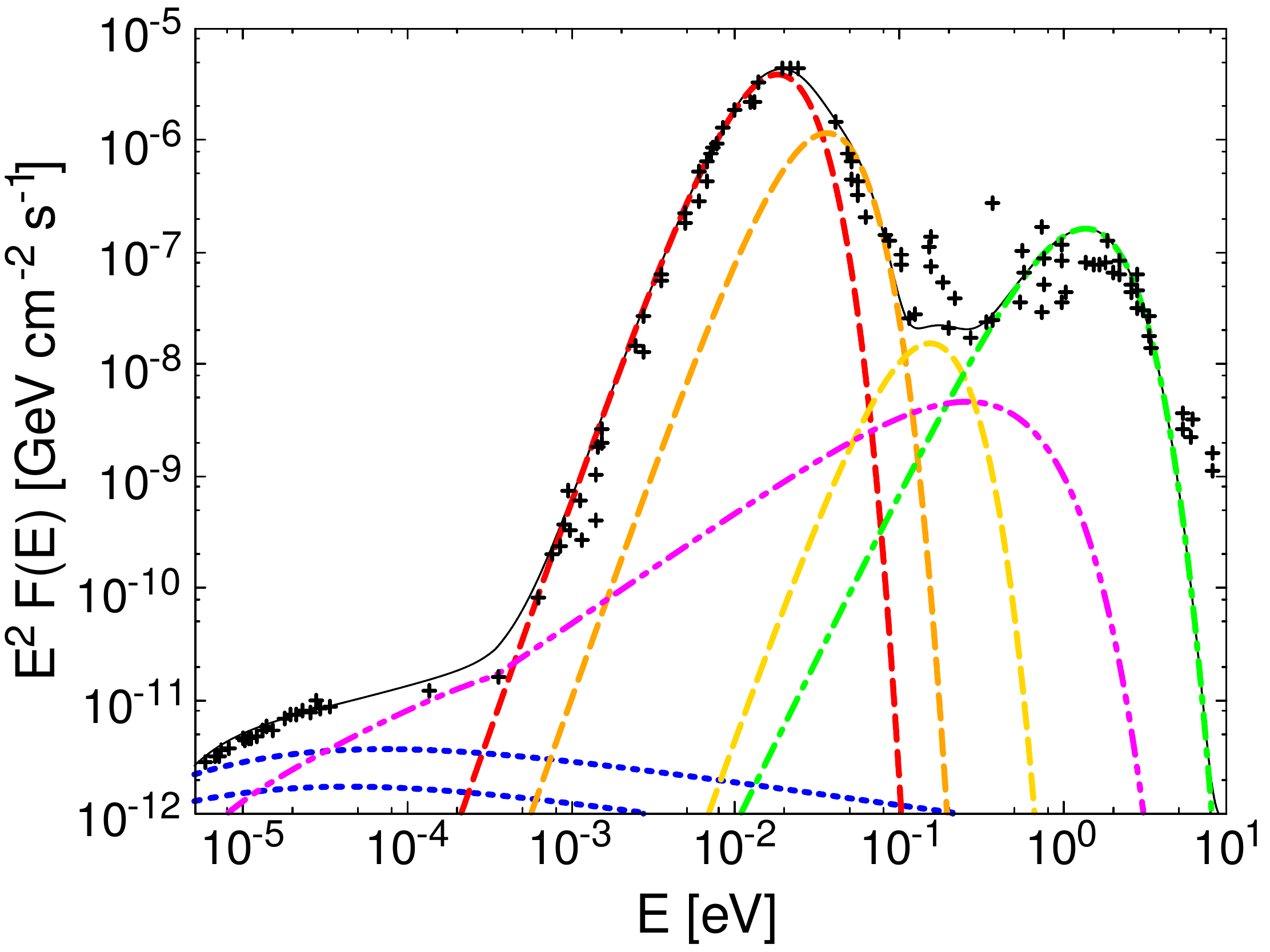}\quad\includegraphics[width=0.45\textwidth]{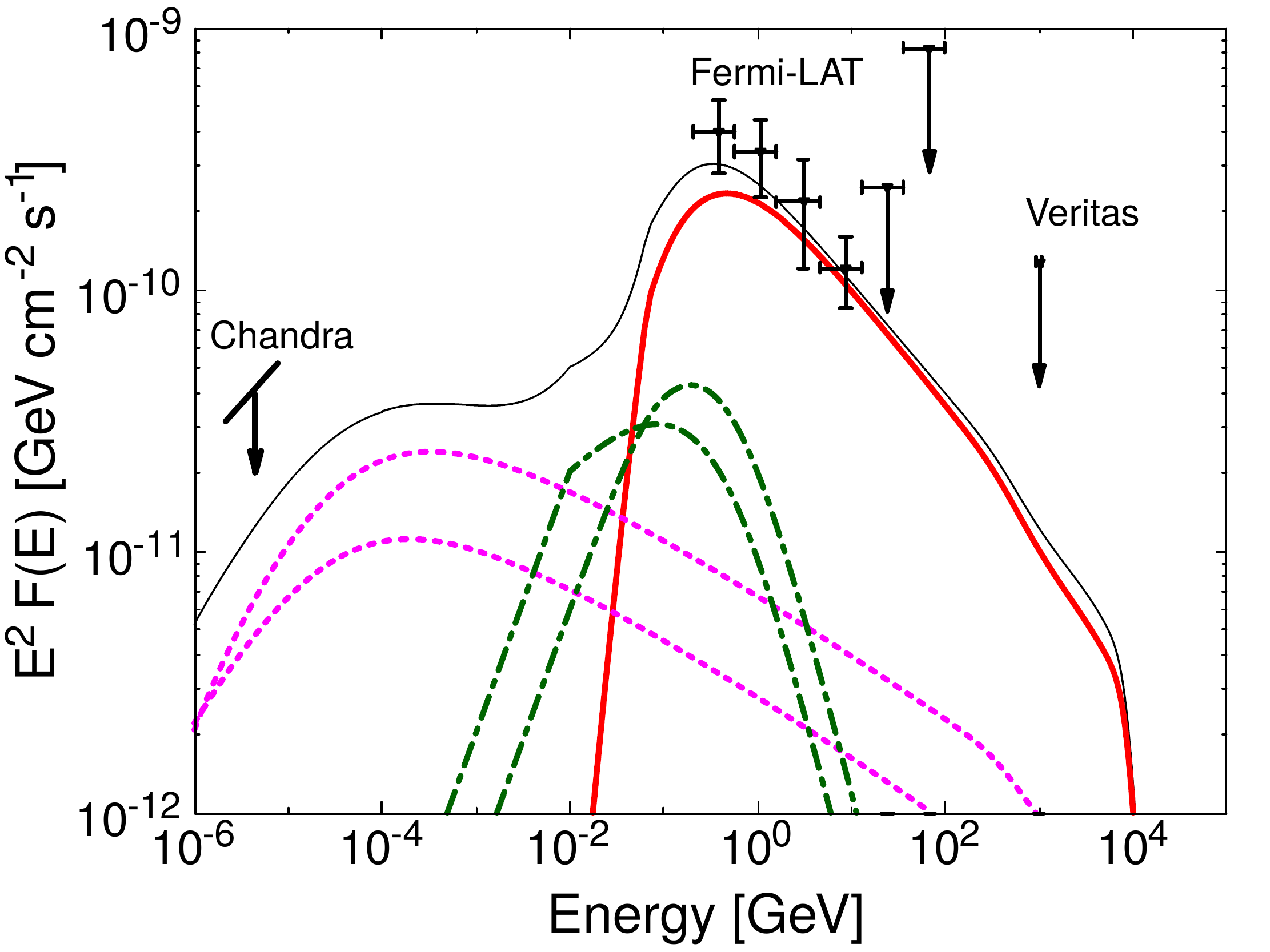}
\caption{\label{fig:ARP220} Multiwavelength spectrum of Arp220. The line stile is the same of figure \ref{fig:NGC253}. Gamma-ray data are taken from \citet{Peng:2016nsx}, whereas the X-ray point (which again we take as upper limit for our diffuse flux taking into account possible contamination from pointlike sources and thermal plasma) has been taken from \citet{2017ApJ...841...44P}.}
\end{figure}

The multifrequency spectrum of Arp220 is shown in Figure \ref{fig:ARP220}. Gamma ray observations \citep[see][]{Peng:2016nsx} suggest that Arp220 requires a softer injection slope with respect to normal starbursts like NGC253 and M82. In alternative, one could speculate that the level of turbulence in Arp220 is lower so as to make CR transport dominated by diffusion. However, this possibility does not seem to sit well with the observed level of activity of this source. On the other hand, it is not easy to envision the reason why one should expect a steeper injection spectrum. In the absence of better indications, here we just assume a steeper injection spectrum. 
 
The dominant gamma-ray component above $\sim 100 \; MeV$ is again the $\pi_0$ decay, whereas at lower energies only ICS and bremsstrahlung emissions are expected to be relevant. Moreover, different from normal starbursts, the synchrotron component is completely negligible in the whole high energy part of the photon spectrum (see lower panel of Figure \ref{fig:ARP220}).

The diffuse hard X-ray flux from the central region of Arp220 has been investigated by \citet{2017ApJ...841...44P}. Taking into account that we are modelling the core of Arp220 as a unique region we show their measured X-ray luminosity coming from the central $4.5''$, corresponding to a radius of $\sim 840$ pc that also accounts for the region between the two nuclei (see X-ray upper limit in the lower panel of Figure \ref{fig:ARP220}). As for the other two SBNi analyzed above, we take this measured luminosity as an upper limit for our non-thermal X-ray flux because of possible contamination from pointlike sources. Indeed, after converting the measured luminosity in a differential flux assuming an energy slope of $-1.6$, we find out that the measured flux is located above our computed spectrum, as expected.

The application of our calculations of CR transport to individual SBGs allows us to draw some general conclusions: 1) in all cases we considered, observations show that CR protons lose an appreciable fraction of their energy inside the SBN; 2) from the point of view electrons, the SBN is an excellent calorimeter. 3) Most of the emission at frequencies other than high energy gamma rays is dominated by secondary electrons, products of pp collisions. 4) Electron positron pairs are effectively generated because of the absorption of high energy gamma rays with the background light in the SBN. The absorption of gamma rays inside the nucleus inhibits the development of an electromagnetic cascade during propagation, which might have important implications for the sources of high energy neutrinos; 5) The synchrotron emission of secondary and tertiary electrons generates a diffuse hard X-ray emission that can be envisioned as a unique diagnostic to investigate the calorimetric properties of SBGs. 


More detailed observations of gamma-ray emission from SBGs with upcoming telescopes, and in particular with the Cherenkov Telescope Array \citep[see][]{2017arXiv170907997C}, will certainly shed new light on the physical processes at work in SBGs.  


\begin{table}
\centering
\begin{tabular}{|c|c|c|c|}
\hline
 Outcomes  &  NGC253  & M82 & Arp220 \\ \hline \hline
 $E_{\rm SN} \xi_{\rm CR}$ (erg s$^{-1}$) & $ 8.56 \times 10^{40}$ & $ 1.59 \times 10^{41}$  & $ 7.14 \times 10^{42}$ \\ \hline
 $L_{0.1-10^2 \rm GeV}$ (erg s$^{-1}$) & $ 1.31 \times 10^{40}$ & $ 1.82 \times 10^{40}$ & $ 1.36 \times 10^{42}$   \\ \hline
  $L_{1-10^2 \rm keV}$ (erg s$^{-1}$) & $ 0.81 \times 10^{39} $ & $ 1.51 \times 10^{39} $ & $ 9.91 \times 10^{40} $   \\ \hline
 $L_{8-10^3 \mu \rm m}$ (erg s$^{-1}$) & $ 1.65 \times 10^{44}$  & $ 2.27 \times 10^{44}$  & $ 6.51 \times 10^{45}$   \\ \hline
 $U_{B}$ (eV cm$^{-3}$) & $  717.71$  & $  1095.19$  & $ 6208.54$   \\ \hline 
 $U_{p}$ (eV cm$^{-3}$) & $ 655.63$  & $ 413.29$  & $ 1323.91$   \\ \hline
 $U_{e}$ (eV cm$^{-3}$) & $ 5.06$ & $ 3.41$ & $ 14.35$  \\ \hline
 $U_{e,\rm sec}$ (eV cm$^{-3}$) & $ 6.15$ & $ 3.92$ & $ 15.78$  \\ \hline
 $U_{e, \rm ter}$ (eV cm$^{-3}$) & $ 3.48 \times 10^{-3}$ & $ 1.49 \times 10^{-3}$ & $ 2.65 \times 10^{-3}$  \\ \hline
\end{tabular}
\caption{\label{tab:outcomes_fit_table} Inferred values for the luminosity at different energies and energy density of magnetic field and non thermal particles for the examined galaxies.}
\label{tab:results}
\end{table}

The single-flavor neutrino fluxes are well described by power laws in energy of index $\alpha - 2$. The flux normalization at $10^2$ TeV obtained for NGC253 and M82 is roughly $10^{-11}$ GeV cm$^{-2}$ s$^{-1}$, and it is about a factor $50$ lower for Arp220. 
Considering that the pointlike source sensitivity for IceCube and KM3NeT allows for the detection of a neutrino flux two orders of magnitude higher than what we obtained for NGC253 and M82 \citep[see][]{Aartsen:2017kru,Aiello:2018usb}, the probability of detecting a nearby SBN as an isolated neutrino source is very small.

\section{Conclusions}
\label{Conclusioni}
We have modeled starburst nuclei as leaky box systems assuming spherical symmetry and homogeneous properties of the medium. We have investigated how different diffusion coefficients change the high energy spectra modifying the normalization and the slope in the energy range above GeV and determining an enhanced flux in the hard X-ray energy band. We have found that in the most likely diffusion scenario, which is described by a Kolmogorov diffusion coefficient assuming $\delta B/B \approx 1$ and typical length of perturbation $L_0 \approx 1$ pc, the escape is completely provided by the wind advection up to PeV energies. At higher energies the timescale at which particles can diffuse away could become comparable with advection and energy losses.

Normal starbursts like NGC253 and M82 are consistent with a slope of injection $\alpha \approx 4.2 \div 4.3$ and the softening taking place in the high energy part of their photon spectra can be explained by the $\gamma \gamma$ absorption. On the other hand the ULIRG class Arp220 is compatible with a softer injection $\alpha = 4.45$. Moreover, in agreement with the results obtained in \cite{Yoast-Hull_Equipartition}, the galaxies we have analysed are consistent with a sub-equipartition between the energy density of CR-particle and the magnetic field, namely $U_p/U_B \approx$ $ 0.9$, $ 0.4$ and $0.2$ for NGC253, M82 and Arp220, respectively.  The ratio between the gamma-ray luminosity and the total injected energy in CRs ($\ge 1/10$) suggests that proton calorimetry is at least partially achieved in NGC253 and M82, whereas Arp220 seems to be able to better confine particles (showing a ratio $\sim 1/5$). 

The neutrino flux from individual SBNi was found to be well below the point source sensitivity of current neutrino telescopes. On the other hand, as pointed out by \citet[]{Loeb:2006tw,2011ApJ...734..107L,Tamborra:2014xia,Bechtol:2015uqb}, the contribution of SBNi to the diffuse neutrino flux might be relevant. The implications of the CR confinement studied in the present paper for the diffuse neutrino flux will be discussed in an upcoming article.

\section*{Acknowledgements}

This research has made use of the NASA/IPAC Extragalactic Database (NED), which is operated by the Jet Propulsion Laboratory, California Institute of Technology, under contract with the National Aeronautics and Space Administration.




\bibliographystyle{mnras}
\bibliography{reference} 





\appendix

\section{Energy loss timescales}
\label{app:timescales}

In this appendix we summarise the formula used for the energy losses of all processes considered for both electrons and protons. The energy loss timescale due to a generic process, $j$, is usually defined from the emitted power as
\begin{equation}
    \tau_{{\rm loss},j} = \left[ \frac{1}{E} \left( \frac{dE}{dt} \right)_{{\rm loss},j} \right]^{-1}
\end{equation}

The electron synchrotron timescale is derived using the classical emitted power formula (\citealp[see e.g.][]{1986rpa..book.....R,2011hea..book.....L})
\begin{equation}
P_{\rm syn} \equiv \left( \frac{dE}{dt} \right)_{\rm syn} = \frac{4}{3} \sigma_T c \gamma^2 \beta^2 U_B.
\end{equation}
The inverse Compton emitted power is \citep[see][]{PhysRev.167.1159}
\begin{align}
\begin{split}
\left( \frac{dE}{dt} \right)_{\rm IC} 
	= \frac{3 \sigma_T c m_e^2 c^4}{4}   
	 \int_{0}^{\infty} d \epsilon \; \frac{n(\epsilon)}{\epsilon} \int_0^1 dq \; \frac{\Gamma^2 q \, G(q, \Gamma)}{(1+ \Gamma q)^3}  \;,
\end{split}
\end{align}
where the function $G(q,\Gamma)$ (which is part of the Klein-Nishina cross section) is defined as 
\begin{equation}
G(q,\Gamma)=2q \log(q)+(1+2q)(1-q)+ (\Gamma q)^2(1-q)/2(1+\Gamma q)
\end{equation}
with $\Gamma=4 \epsilon E_e/m_e^2c^4$ and $q=E/\Gamma/(E_e-E)$. 
For the bremsstrahlung timescale we used \citep[see e.g.][]{2013SAAS...40.....A}
\begin{equation}
\tau_{\rm brem} \approx 4 \times 10^{7} \left( n/{\rm cm^{-3}} \right)^{-1} \,{\rm yr} \,,
\end{equation}
whereas, the ionization timescale for electrons is given by \cite[see][]{2002cra..book.....S}
\begin{equation}
\tau_{\rm ion} \approx 1.9 \times 10^9  \; \left( \frac{E}{100 \, \rm TeV}  \right) \left( \frac{n}{250 \, \rm cm^{-3}}  \right)^{-1} \,{\rm yr},
\end{equation}
where we approximated the  $\ln(\gamma)$ term to its value at $10$ TeV as also done by \cite{0004-637X-762-1-29}.

Concerning protons, the timescale for proton-proton inelastic scattering is given by
\begin{equation}
\tau_{\rm pp}= \left(n \sigma_{pp}(E)c/\kappa \right)^{-1},
\end{equation}
where $\kappa$ is the inelasticity of the process and is  $\kappa \approx 3 K_{\pi} \approx 0.5$, where we used the value of $K_\pi$ below Eq.~(\ref{pion_injection}).
For the proton ionization energy loss we use the following expression form \cite{2002cra..book.....S}
\begin{align} 
\begin{split}
	\left( \frac{dE}{dt} \right)_{\rm ion} \approx 1.82 \times 10^{-7} n \; \times \hspace{4cm}\\ 
		\left[1+0.0185 \ln(\beta) \, \Theta(\beta-0.01) \right] \frac{2 \beta^2}		
		{10^{-6}+2 \beta^3} \; \rm eV \, s^{-1} \,,
\end{split}
\end{align}
where $\beta = v/c$, and $\Theta$ is the step function. The Coulomb energy loss can be approximated by \citep[see again][]{2002cra..book.....S}
\begin{align}
 \begin{split} 
 \left( \frac{dE}{dt} \right)_{\rm Coul} \approx  \frac{3.08 \times 10^{-7} 
 		n_e \beta^2}{\beta^3+2.34 \; 10^{-5}(T_e/2.0 \; 10^6K)^{1.5}}  \; \times \hspace{1cm} \\ 
 	  \Theta \left[ \beta -7.4 \times 10^{-4} (T_e/2.0 \; 10^6K)^{1/2} \right] \; \rm eV \, s^{-1} \,,
\end{split}
\end{align}
where and $T_e$ and $n_e$ are the plasma temperature and density, respectively.

\section{Production of secondary particles} \label{app:secondaries}
For the sake of completeness, here we report the detailed formula used to calculate the production of secondaries as discussed in details in \cite{2006PhRvD..74c4018K}.
In the following equations we use the parameters $x \equiv E_j/E_{\pi}$ and $r \equiv 1-\lambda=(m_{\mu}/m_{\pi})$, where the former is the fraction of energy of the lepton "$j$" with respect the parent $\pi$ meson, whereas the latter is the ratio between the muon and pion masses.

The muonic neutrinos produced in the direct process $\pi \longrightarrow \mu \nu_{\mu}$ is described by
\begin{equation}
f_{\nu_{\mu}^{(1)}}(x)= \frac{1}{\lambda} \; \delta(E_{\nu_e}-E_{\nu_{\mu}}) \; \theta[\lambda - x] \,,
\end{equation}
while the electrons and muonic neutrinos produced by the muon decay $\mu \longrightarrow \nu_{\mu} \bar{\nu}_{e}e$ are described by the following function
\begin{align}
\begin{split}
f_e (x) = f_{\nu_{\mu}^{(2)}} (x) = g_{\nu_{\mu}}(x) \; \theta[x-r] + \\ + \big[  h^{(1)}_{\nu_{\mu}}(x) + h^{(2)}_{\nu_{\mu}}(x) \big] \; \theta[r-x] 
\end{split}
\end{align}
where the function $g$ and $h$ are defined as
\begin{align*}
    g_{\nu_{\mu}}(x)= (9x^2 - 6 \ln(x)-4x^3-5) (3-2r)/9/(1-r)^2 \\ h^{(1)}_{\nu_{\mu}}(x)= (9r^2 - 6 \ln(r)-4r^3-5) (3-2r)/9/(1-r)^2 \\
    h^{(2)}_{\nu_{\mu}}(x)= [9(r+x)- 4 (r^2 + r x +x^2)]   (1+2r)(r-x)/(9r^2)
\end{align*}

The electron neutrino function is
\begin{equation}
f_{\nu_{e}} (x) = g_{\nu_{e}}(x) \; \theta[x-r] +  \big[  h^{(1)}_{\nu_{e}}(x) + h^{(2)}_{\nu_{e}}(x) \big] \; \theta[r-x]
\end{equation}
where the functions $g$ and $h$ are defined as
\begin{align*}
g_{\nu_{e}}(x)=\frac{2(1-x) \big[ 6(1-x)^2 +r(5+5x-4x^2)+6r \ln(x) \big]}{3 (1-r)^2} \\ 
h^{(1)}_{\nu_{e}}(x)= \frac{2 \big[ (1-r)(6-7r+11r^2-4r^3+6r \ln(r) \big]}{3(1-r)^2} \\
h^{(2)}_{\nu_{e}}(x)= \frac{2(r-x) \big( 7r^2 - 4r^3 + 7xr - 4xr^2 - 2x^2 - 4x^2r \big)}{3 r^2}.
\end{align*}

\section{EBL approximation} \label{app:EBL}
In this work we use the EBL model developed in \cite{2017A&A...603A..34F} to account for $\gamma \gamma$ absorption during the propagation in the intergalactic medium. From a fitting procedure, we found the following analytic approximation which is able to reproduce their result for $z=0.003$ with an accuracy $\lesssim 6 \%$. The approximate expression reads:
\begin{align}
\begin{split}
\tau_{\gamma \gamma}^{*}(E) = \frac{95}{1100}   \; \Bigg\{  \Big[ \frac{(E/1 \; TeV)^{-2.7}}{2.1} + \frac{(E/ 1 \; TeV)^{-0.31}}{0.34}  \Big]^{-1}  +  \; \\ \;
\Big[ \frac{(E/12 \; TeV)^{-3.1}}{0.47} + \frac{(E/ 40 \; TeV)^{-0.8}}{20}  \Big]^{-1} + 7 \Big( \frac{E}{100 \; TeV} \Big)^{7.8}
\Bigg\} .
\end{split}
\end{align} 
We introduced the redshift dependence in our analytic formula of the optical depth as follows
\begin{equation}
    \tau_{\gamma \gamma}(E,z)= \tau_{\gamma \gamma}^{*}(E) \Big[ \frac{z}{0.003} \Big],
\end{equation}
which has an accuracy $\lesssim 10 \%$ when $z=0.01$ and $\lesssim 20 \%$ when $z=0.03$.

\section{Analytic estimates for  production of secondary and tertiary electrons}
\label{Appendice_Stime_Analitiche}
In this Appendix, using simple analytical estimates, we show that secondary and tertiary electrons are always as important as primary ones if the SBN behaves approximately as a calorimeter for CR protons. We assumed throughout the paper that the injection spectrum of primary electrons is  related to the protons one as $q_e = q_p(E)/50$. On the other hand, the injection of secondary electrons from $pp$ scattering can be written as
\begin{equation}
	q_{pp\rightarrow e}(E) = n_{\rm ISM} \, \sigma_{pp} \, c \, f_p(E/\xi_{e}) /\xi_{e}  \,,
\label{eq:sec_el}
\end{equation}
where $\xi_{e}\approx 0.05$ is the fraction of parent proton's energy transferred to the electron. The parent proton's spectrum is $f_p(E)= q_p(e) \tau_{\rm loss}(E)/\eta$, where the factor $\eta$ accounts approximately for the transport condition: $\eta= 1$ for the calorimetric or 2 for the escape limited scenarios, respectively. Now, using $q_p(E) \propto E^{-\alpha+2}$ and $\tau_{\rm loss} = (n_{\rm ISM} \sigma_{pp} c)^{-1}$, and neglecting the mild energy dependence of $\sigma_{pp}$, one easily found that 
\begin{equation}
	\frac{q_{pp\rightarrow e}(E)}{q_e(E)} \simeq \frac{50}{\eta} \xi^{\alpha-3} \approx 1
\end{equation}
where we used $\alpha=4.3$.

A similar result is valid for tertiary electrons produced from pair production. In this case we can use the leading particle approach, where the energy of the photons is assumed to be transferred only to one of the two electrons. With this approach we show that the ratio between tertiary and secondary electrons is approximatively given by the $\gamma \gamma$ optical depth. 
The injection term for electron due to pair production reads
\begin{align}
\begin{split}
q_{\gamma\gamma \rightarrow e}(E)&= \int d\epsilon \,  n_{\rm bkg}(\epsilon) n_{\gamma}(E) \sigma_{\gamma \gamma}(E,\epsilon) \, c \\ 
		  &=  n_{\gamma}(E)c \tau_{\gamma \gamma}(E)/R.
\end{split}
\label{eq:ter_el}
\end{align}
Taking into account that the high energy photon density can be approximated by the $\gamma$-rays injected via the dominant $\pi_0$ decay mechanism, the following relation can be assumed
\begin{equation}
   n_{\gamma}(E) = \frac{R}{c} q_{\pi_0 \rightarrow 2 \gamma}(E)
			   = \frac{R}{c} n_{\rm ISM} \sigma_{pp} f_p(E/ \xi_{\gamma}) / \xi_{\gamma}\,
\label{eq:n_gamma}
\end{equation}
where $\xi_\gamma$ is the proton's energy transferred to the photon.
Now, using Eqs.~(\ref{eq:ter_el}), (\ref{eq:n_gamma}) and (\ref{eq:sec_el}) we can estimate the ratio between pair production and $pp$ electrons as follows
\begin{equation}
   \frac{q_{\gamma\gamma \rightarrow e}}{q_{pp \rightarrow e}} 
   			 \approx \frac{(R/c) \, n_{\rm ISM} \sigma_{pp}  f_p(E/\xi_{\gamma})/\xi_{\gamma} \, c \, \tau_{\gamma \gamma}/R}
				{n_{\rm ISM} \sigma_{pp} \, c\, f_p(E/\xi_{e})/\xi_{e}} 
			 = \tau_{\gamma \gamma} \,,
\end{equation}
where we assumed $\xi_{e} \approx \xi_{\gamma}$.
The later equality is valid only when $\tau_{\gamma \gamma} \le 1$ then it saturates giving $q_{\gamma\gamma \rightarrow e} / q_{pp \rightarrow e} = 1$.
Since the $\gamma \gamma$ optical depth can be roughly estimated using only the IR photon density, i.e.
\begin{equation}
\begin{split}
   \tau_{\gamma \gamma} (100 \, {\rm TeV}) &\approx \, 0.2 \sigma_T \, R \, n_{\rm ph}(0.01 \, {\rm eV}) \\ 
   		& \approx 4.123 \, \left( \frac{R}{100 \rm pc} \right) \left( \frac{U_{\rm rad}}{10^3 \, \rm eV cm^{-3}} \right) 
		\left( \frac{\epsilon}{0.01 \, \rm eV} \right)^{-1}
\end{split}
\end{equation}
we observe that, for standard physical condition in a SBN, $\tau_{\gamma\gamma}$ easily reaches values larger than $1$. Therefore, for a large energy range, one may expect that tertiary electrons are as important as secondary one.

A further channel for the production of secondary electrons is through the $p\gamma$ interaction. Nevertheless, the energy threshold of protons for such a process is $\sim 3 \times 10^{17}$ eV for protons interacting with $1$ eV background photons and two orders of magnitude higher for the interaction with the FIR background. 
Considering that SBNi are probably incapable to accelerate protons up to such a very high energies, neglecting this term, as we did in this work, is widely justified.

\bsp	
\label{lastpage}
\end{document}